\documentclass[]{aa}       % New LaTeX A&A Standard Fonts
\usepackage{graphicx}
\usepackage{amssymb}
\usepackage{longtable}
\usepackage{supertabular}
\usepackage{natbib} 
\usepackage{ulem}

\begin{document}

\title{Orbital evolution of colliding star and pulsar winds in 2D and 3D;\\
effects of: dimensionality, EoS, resolution, and grid size}

\author{V. Bosch-Ramon\inst{1},
M.V. Barkov \inst{2}, \and
M. Perucho\inst{3}
}

\authorrunning{Bosch-Ramon, Barkov \& Perucho}

\titlerunning{Colliding star and pulsar winds in 2D and 3D}

\institute{
Departament d'Astronomia i Meteorologia, Institut de Ci\`encies del Cosmos (ICC), Universitat de Barcelona (IEEC-UB), Mart\'i i
Franqu\`es 1, 08028 Barcelona, Catalonia, Spain
\and
Astrophysical Big Bang Laboratory, RIKEN, 2-1 Hirosawa, Wako, Saitama 351-0198, Japan
\and
Dept. d'Astronomia i Astrof\'{\i}sica, Universitat de Val\`encia, C/ Dr. Moliner 50, 46100, Burjassot (Val\`encia), Spain
}

\offprints{\email{vbosch@am.ub.es}}

\date{Received <date> / Accepted <date>}

\abstract
{The structure formed by the shocked winds of a massive star and a non-accreting pulsar in a binary system suffers periodic and random variations of orbital and non-linear dynamical origins. The characterization of the evolution of the wind interaction region is necessary for understanding the rich phenomenology of these sources.}
{For the first time, we simulate in 3 dimensions the interaction of isotropic stellar and relativistic pulsar winds along one full orbit, on scales well beyond the binary size. We also investigate the impact of grid resolution and size, and of different state equations: a $\hat\gamma$-constant ideal gas, and an ideal gas with $\hat\gamma$ dependent on temperature.}
{We used the code {\it PLUTO}{} to carry out relativistic hydrodynamical simulations in 2 and 3 dimensions of the interaction between a slow dense wind and a mildly relativistic wind with Lorentz factor 2, along one full orbit in a region up to $\sim 100$ times the binary size. The different 2-dimensional simulations were carried out with equal and larger grid resolution and size, and one was done with a more realistic equation of state than in 3 dimensions.}
{The simulations in 3 dimensions confirm previous results in 2 dimensions, showing: a strong shock induced by Coriolis forces that terminates the pulsar wind, closing it in all directions; strong bending of the shocked-wind structure against the pulsar motion; and the generation of turbulence. The shocked flows are also subject to a faster development of instabilities in 3 dimensions, which enhances the presence of shocks, two-wind mixing, and large-scale disruption of the shocked structure. In 2 dimensions, higher resolution simulations confirm lower resolution results, simulations with larger grid sizes strengthen the case for the loss of the general coherence of the shocked structure, and simulations with two different equations of state yield very similar results. In addition to the Kelvin-Helmholtz instability, discussed in the past, we find that the Richtmyer-Meshkov and the Rayleigh-Taylor instabilities are very likely acting together in the shocked flow evolution.}
{Simulations in 3 dimensions confirm that the interaction of stellar and pulsar winds yields structures that evolve non-linearly and become strongly entangled. The evolution is accompanied by strong kinetic energy dissipation, rapid changes in flow orientation and speed, and turbulent motion. The results of this work strengthen the case for the loss of the coherence of the whole shocked structure on large scales, although simulations of more realistic pulsar wind speeds are needed.}

\maketitle

\section{Introduction} \label{intro}

% %fffffffffffffffffffffffffffffffffffffffffffffffffffffffffffffffffff
 \begin{figure}[htp]
 \includegraphics[width=80mm]{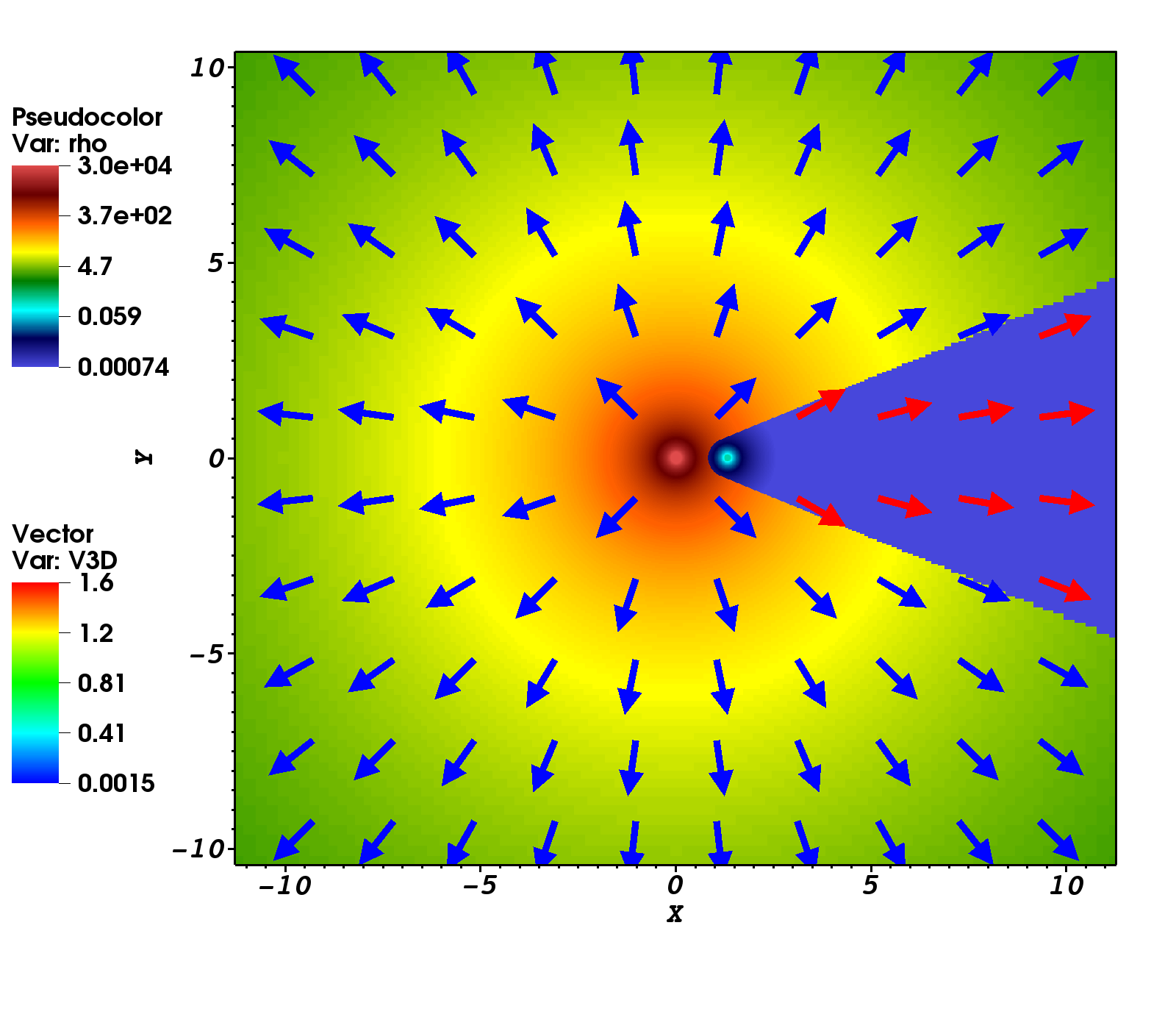}
 \caption{Density distribution by colour and arrows representing the flow motion direction, in the orbital plane (XY) for 3Dlf at the beginning of the simulation ($t=0$; apastron).}
\label{setup}
\end{figure}
% %fffffffffffffffffffffffffffffffffffffffffffffffffffffffffffffffffff

As already proposed long ago, binary systems hosting a massive star and a powerful non-accreting pulsar can produce gamma rays through the interaction of the stellar 
and the pulsar winds \citep[e.g.][]{Maraschi1981}. An actual instance of this kind of object is the binary system PSR~B1259$-$63(/LS2883), which consists of a 
late O star with a decretion disc \citep{neg11} and a 47~ms pulsar \citep{joh92}, and which is a powerful GeV and TeV emitter \citep[e.g.][]{aha05,abd11,tam11}. 
Other binaries emitting gamma rays may also pertain to this class, such as LS~5039, LS~I~+61~303, HESS~J0632$+$057, and 1FGL~J1018.6$-$5856 
\citep[see e.g.][and references therein]{barkha12,Paredes2013,Dubus2013}, although the true nature of the compact object in these sources is still unknown. 
In the whole Galaxy, if taking a lifetime for the non-accreting pulsar of a few times $10^5$~yr (the age of PSR~B1259$-$63) 
and the birth rates for high-mass binaries hosting a neutron star \citep[e.g.][]{1999MNRAS.309...26P}, one might expect $\sim 100$ of these systems \citep[see also][]{Paredes2013,Dubus2013}.

%fffffffffffffffffffffffffffffffffffffffffffffffffffffffffffffffffff
\begin{figure*}[htp]
\includegraphics[width=160mm]{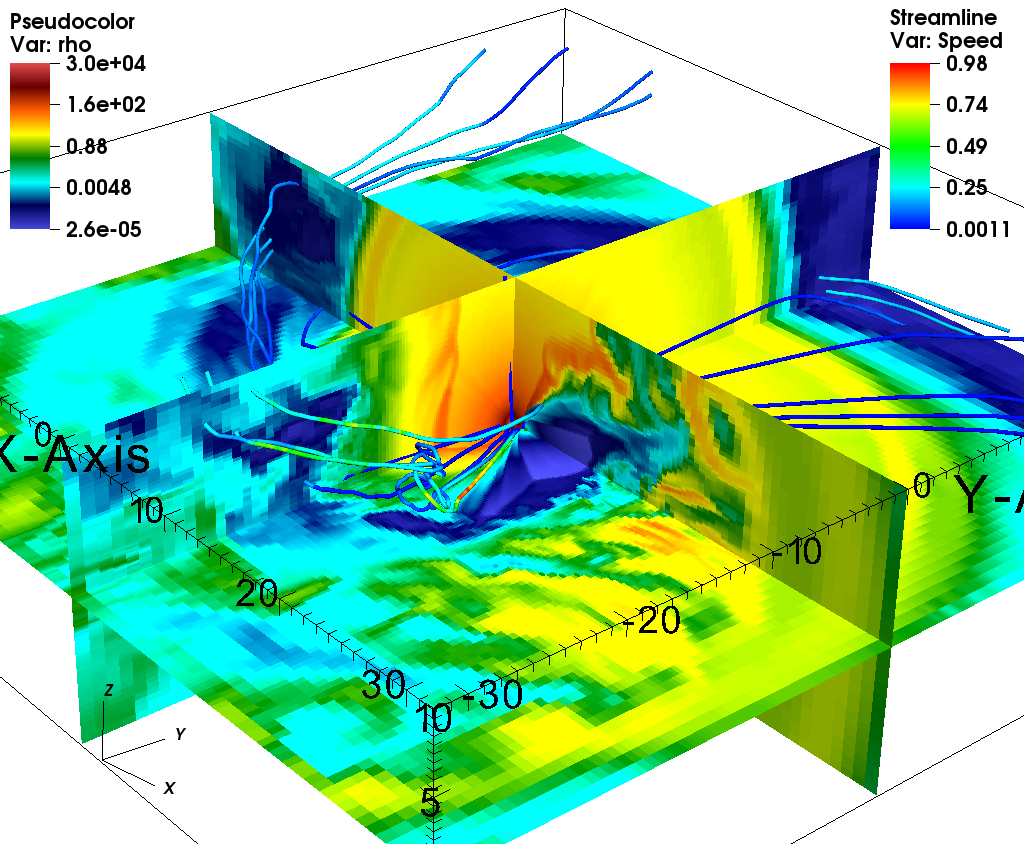}
\caption{Representation of the distribution of density in the $XY$-, $XZ$-, and $YZ$-planes for 3Dlf at $t=3.9$~days (apastron). Streamlines are shown in 3D.}
% ti3 model 
\label{fig:3dsl}
\end{figure*}
%fffffffffffffffffffffffffffffffffffffffffffffffffffffffffffffffffff

The rich phenomenology of PSR~B1259$-$63/LS288 \citep[see][for a multi-wavelength study of the 2010 periastron passage]{Chernyakova2014}, and (possibly) also of other gamma-ray binaries, is directly linked to the complex processes taking place in the colliding wind region. To study this complexity, several works have already been devoted to the numerical study of the wind collision region, neglecting or including orbital motion, and on scales similar or beyond those of the binary \citep{Romero2007,bog08,bog12,oka11,Takata2012,bbkp12,bog12,LambertsAAA,LambertsBBB,Lamberts2013,Paredes-Fortuny2015}. In particular, in \cite{bbkp12} the authors performed two-dimensional (2D) simulations in planar coordinates of the interaction of a dense, non-relativistic stellar wind and a powerful, mildly relativistic pulsar wind along one full orbit, up to scales $\approx 40$ times the orbital separation distance (here orbit semi-major axis $a$). These simulations showed the early stages of the spiral structure also found in non-relativistic simulations \citep{LambertsAAA}, but the fast growth of instabilities \cite[see also][]{Lamberts2013} and strong two-wind mixing have suggested the eventual disruption of the spiral and the isotropization of the mass, momentum, and energy fluxes \citep[as predicted in][]{bb11}. 

%fffffffffffffffffffffffffffffffffffffffffffffffffffffffffffffffffff
\begin{figure*}[htp]
\centering
\includegraphics[width=80mm]{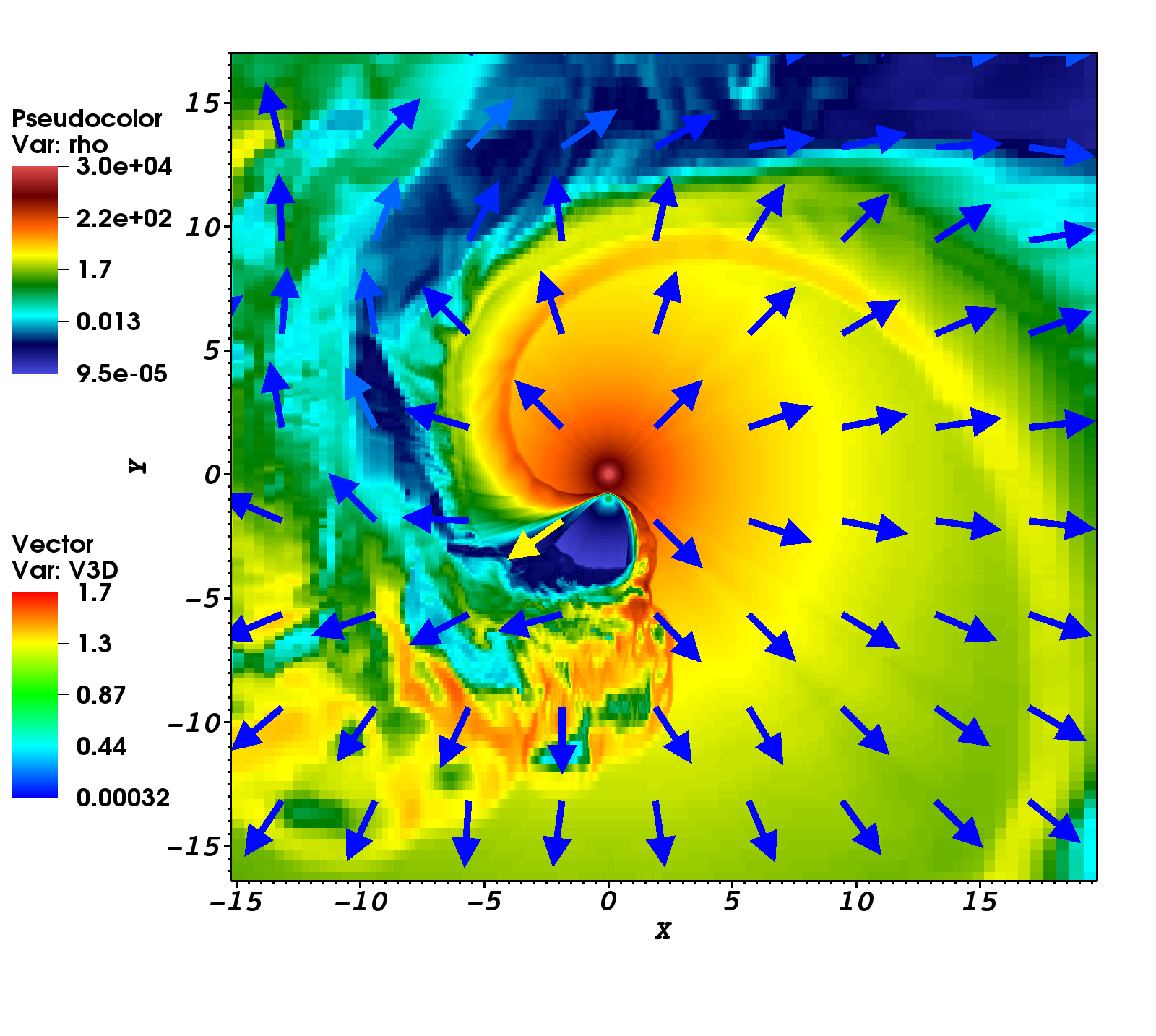}
\includegraphics[width=80mm]{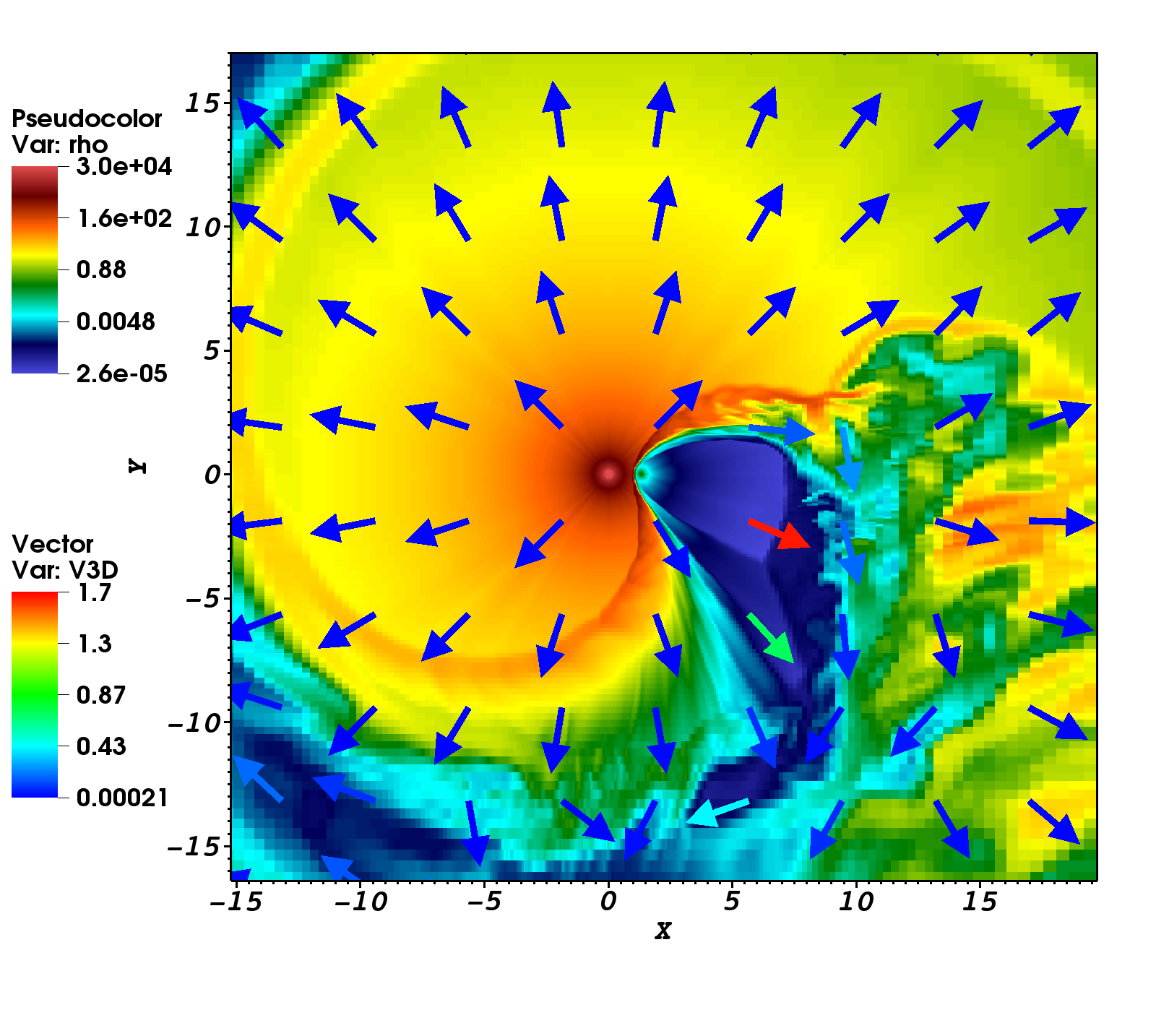}
\includegraphics[width=80mm]{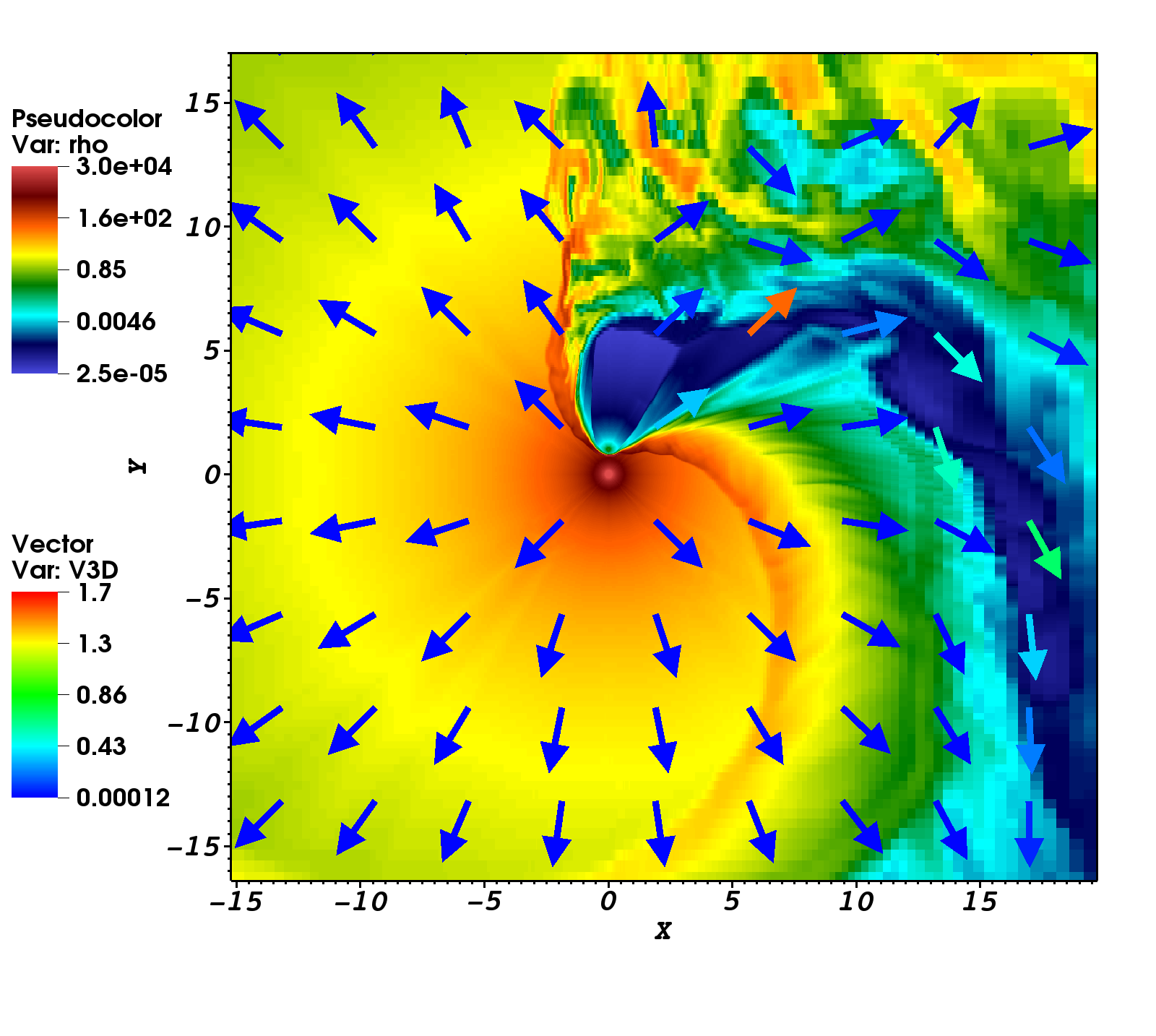}
\includegraphics[width=80mm]{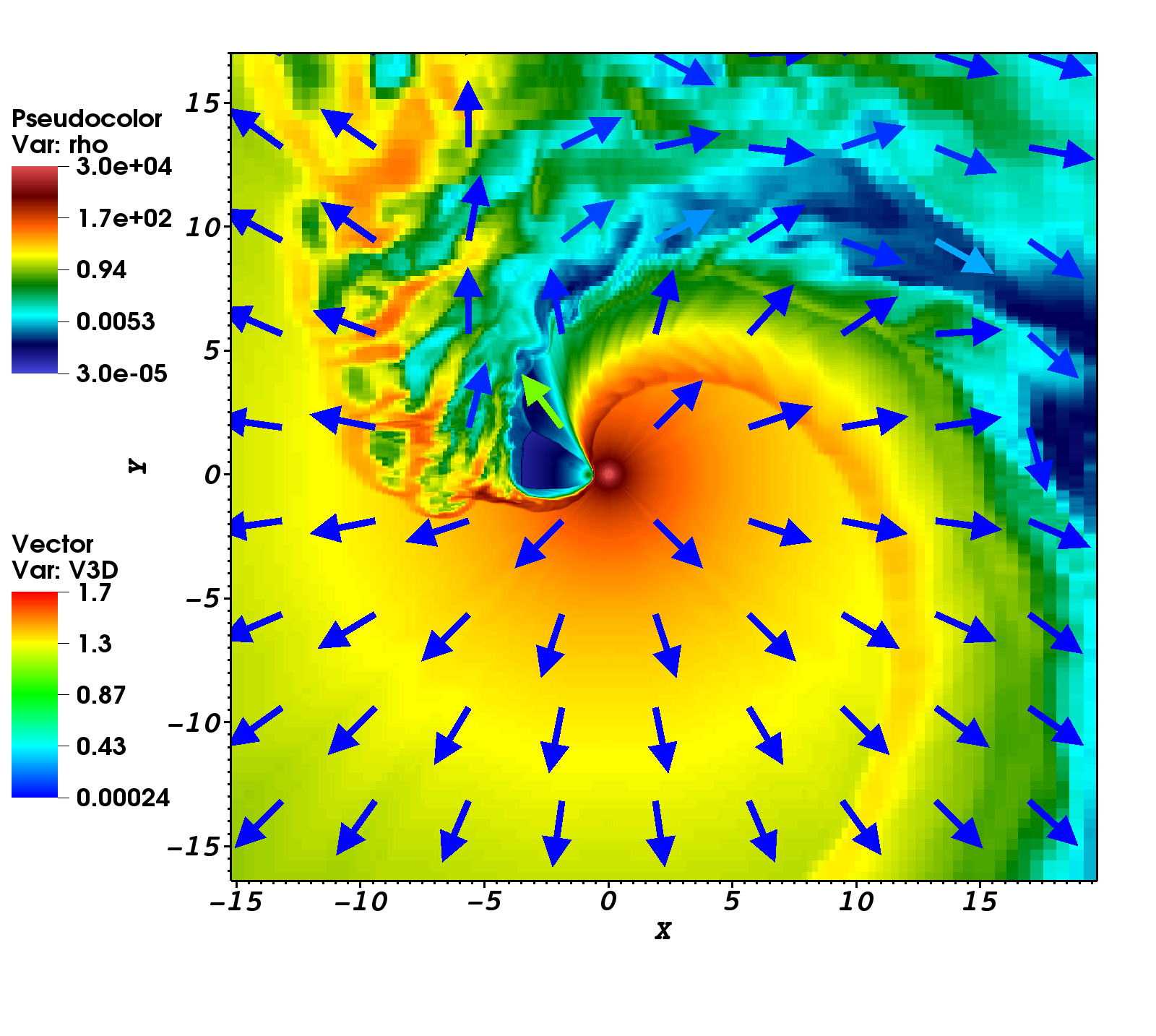}
\caption{Density distribution by colour, and arrows representing the flow motion direction, in the orbital plane (XY) for 3Dlf at: $t=2.6$, 3.9 (apastron), 5.2, and 5.85~days (periastron) (from left to right, and top to bottom).}
% ti3 model 
\label{fig:3Dlfxyoph}
\end{figure*}
%fffffffffffffffffffffffffffffffffffffffffffffffffffffffffffffffffff

If the results from \cite{bbkp12} were confirmed, it would imply several important facts: the two-wind interaction is subject to strongly non-linear processes already within the binary system; the isotropization of the interaction region leads to loss of coherence on scales not far beyond $a$, eventually becoming a more or less irregular isotropic flow formed by mixed stellar and pulsar winds; this mixed flow terminates with a shock on the external medium (interstellar medium -ISM- or the interior of a supernova remnant -SNR-; see \citealt{bb11}). All these dynamical processes would imply the existence of many potential sites for particle acceleration, such as shocks, turbulence, shear flows, and non-thermal emission, mainly synchrotron and inverse Compton, from deep inside the system on pc scales.

In this work, we go a step forward in the study of the interaction of the stellar and the pulsar winds to include orbital motion. We have performed the first three-dimensional (3D), 
relativistic simulations using similar parameters to those adopted in \cite{bbkp12} up to a distance $\sim 30\,a$ from the binary. The results obtained in 3D are compared with 
results found in new 2D simulations in planar coordinates under the same conditions, in particular the same grid size and number of cells in each direction. Additional 2D simulations 
were also carried out to study the effects of a more realistic equation of state (EoS), higher resolution, and a larger grid size up to $\sim 100\,a$ from the binary, not reached so far in relativistic simulations. 

\section{Numerical simulations}\label{num}

%fffffffffffffffffffffffffffffffffffffffffffffffffffffffffffffffffff
\begin{figure}[htp]
\includegraphics[width=80mm]{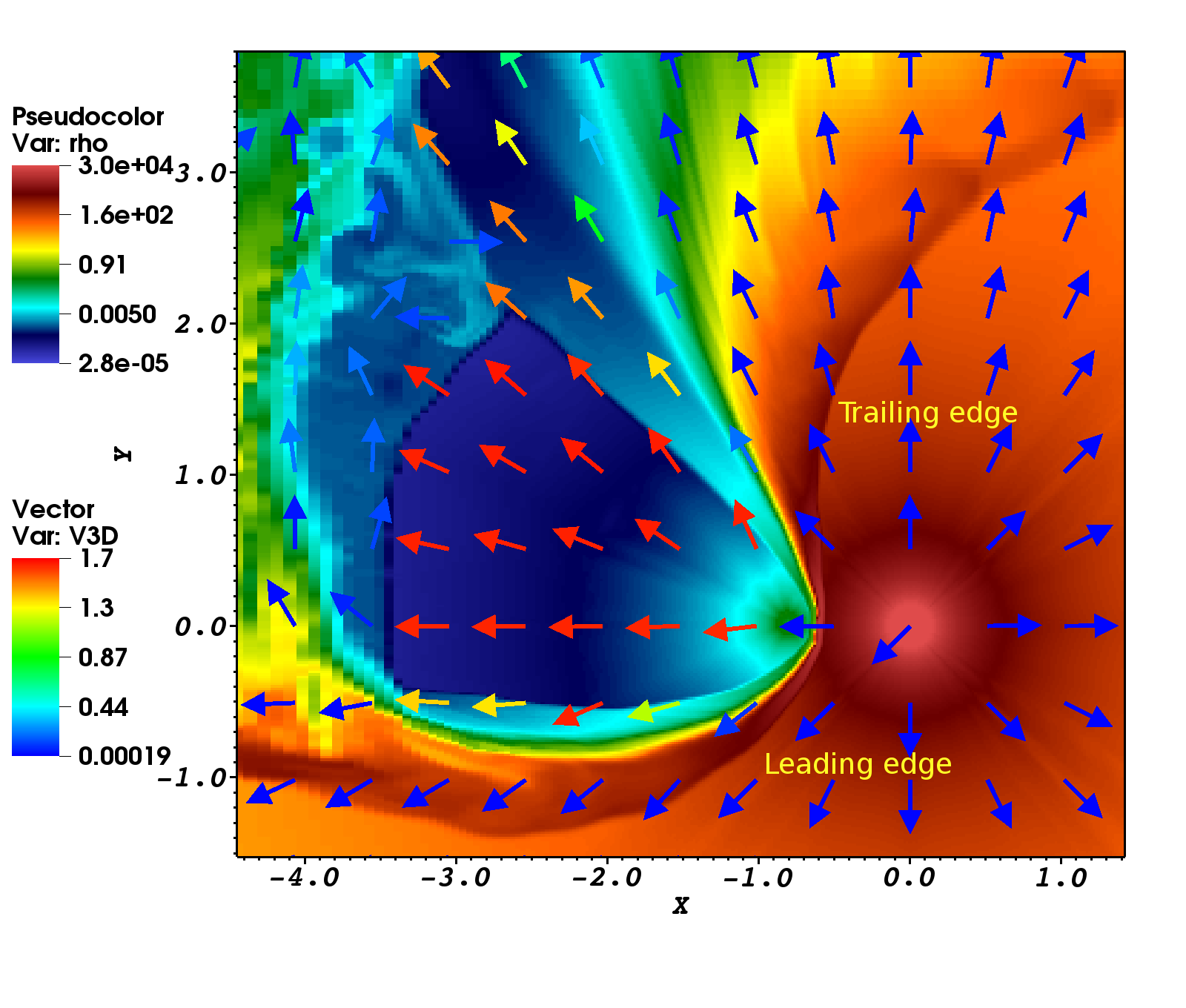}
\caption{Zoom in of the region around the pulsar region at periastron shown in Fig.~\ref{fig:3Dlfxyoph}. Two labels pointing to the two edges, leading and trailing, of the
shocked structure are shown.}
% ti3 model 
\label{fig:3Dzoom}
\end{figure}
%fffffffffffffffffffffffffffffffffffffffffffffffffffffffffffffffffff

\subsection{Simulation code: PLUTO}

The simulations were implemented in 3D and 2D with the {\it PLUTO} code\footnote{Link http://plutocode.ph.unito.it/index.html} \citep{mbm07}, the piece-parabolic method (PPM) \citep{cw84}, and an HLLC Riemann solver \citep{mig05}. {\it PLUTO} is a modular Godunov-type code entirely written in C and intended mainly for astrophysical applications and high Mach number flows in multiple spatial dimensions. For this work, it was run through the MPI library in the CFCA cluster of the National Astronomical Observatory of Japan.

\subsection{Gas equation of state}

The simulated flows were approximated in most of the cases as an ideal, relativistic adiabatic gas with no magnetic field, one particle
species, and a constant polytropic index of 4/3 (CtGa), since using such a simple EoS reduces the time cost of the simulations carried out. One 2D simulation (model 2Dlf; see Table~\ref{tab:models}) was also carried out with an adaptive Synge-type EoS \citep{taub78} for a one particle species, non-degenerate, relativistic gas \citep{mpb05}, to check the impact of a more complex EoS on the general structure. A generalized, more complete treatment of the EoS leads to changes in the Rankine-Hugoniot conditions and gas thermodynamical properties, in particular for the stellar wind. Because the latter is dense and cold enough, thermal cooling could also be important (see Sect.~\ref{disc}). This more complete treatment of the gas is left for future work.

\subsection{Resolution and grid size}

We adopted a changing resolution in the 3D simulations (models 3Dlf and 3Dls; see Table~\ref{tab:models}) to reach larger scales. The computational domain size is $x \in [-32\,a,32\,a]$,  $y \in [-32\,a,32\,a]$, and  $z \in [-10\,a,10\,a]$. The 3D domain has a resolution of $512\times512\times256$ cells.  
The central part of this domain, $x \in [-2\,a,2\,a]$, $y \in [-2\,a,2\,a]$, and $z \in [-a,a]$, has
a resolution of $256\times256\times128$ cells, and outside this domain the cell size grows exponentially\footnote{Link http://plutocode.ph.unito.it/files/userguide.pdf} 
with distance outwards in each direction. The equivalent uniform resolution for the whole domain would be $4096\times4096\times2560$ cells. The adopted resolution came from a compromise between computational costs and a maximum Lorentz factor. The 3D simulations required about $3\times 10^5$~cpu~h on NAOJ cluster XC30. To achieve a Lorentz factor $\approx 10$, we should have doubled the grid resolution \cite[see Sect.~3.1 in][]{bbkp12}, in particular in the central region, leading to a much longer computation (see Sect.~\ref{disc}). Adaptive mesh refinement (AMR) was not implemented for these simulations, because in such a problem, the AMR technique fills most of the computational volume with the smallest scale grid owing to the growth of turbulence. Nested-grid AMR could avoid such behaviour, but abrupt changes in resolution can produce unphysical pressure jumps. Nevertheless, the chosen geometry in our calculations, with an exponential growth of the grid cells outwards, naturally follows the scales of the problem, which smoothly grow outwards with respect to the system.

\subsection{Simulation models}

We performed four additional 2D simulations to compare the influence of
equation of state, geometry dimensionality, grid resolution, and size. Model
2Dlf has the same domain size and resolution distribution as 3Dlf in the X and
Y directions. Model 2Dle has the same domain size and resolution distribution
as 2Dlf in the X and Y directions, but makes use of Taub. Model 2Dhf has the
same domain size but twice the resolution of 2Dlf. Finally, model 2Dhbf has an
increased computational domain of $x \in [-100\,a,100\,a]$ and $y \in
[-100\,a,100\,a]$ (as before, central regions are $x \in [-2\,a,2\,a]$ and $y
\in [-2\,a,2\,a]$), and a better effective resolution with a cell distribution:
$N_x=[512,512,512]$ and $N_y=[512,512,512]$ cells (equivalent uniform
resolution: $25600\times25600$), where the first number shows the resolution in
the region with negative coordinates, the second one is the resolution in the
uniform central region, and the third one is the resolution in the region with
positive coordinates. The model details are summarized in
Table~\ref{tab:models}.

%fffffffffffffffffffffffffffffffffffffffffffffffffffffffffffffffffff
\begin{figure}[htp]
\centering
\includegraphics[width=70mm]{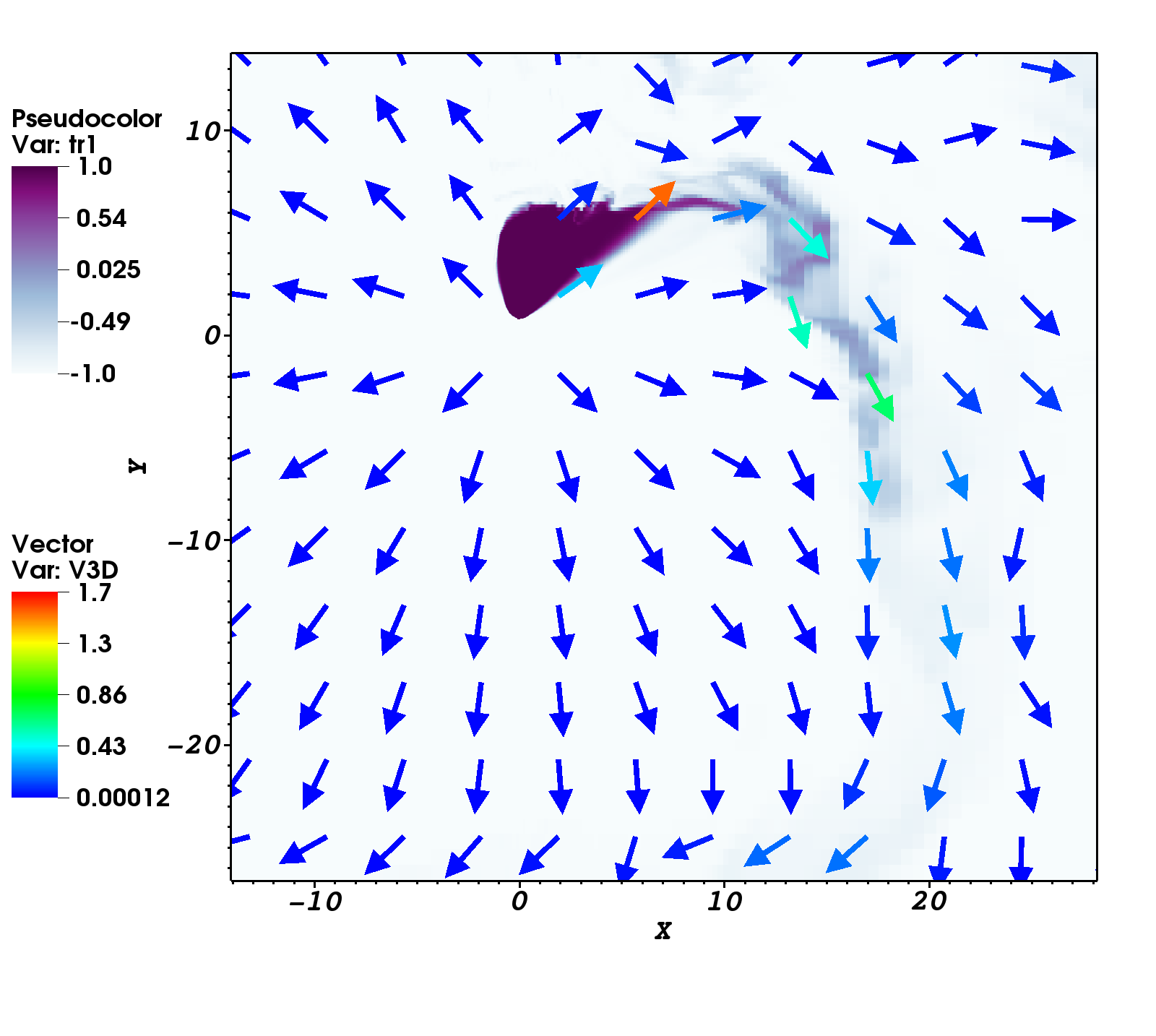}
\includegraphics[width=70mm]{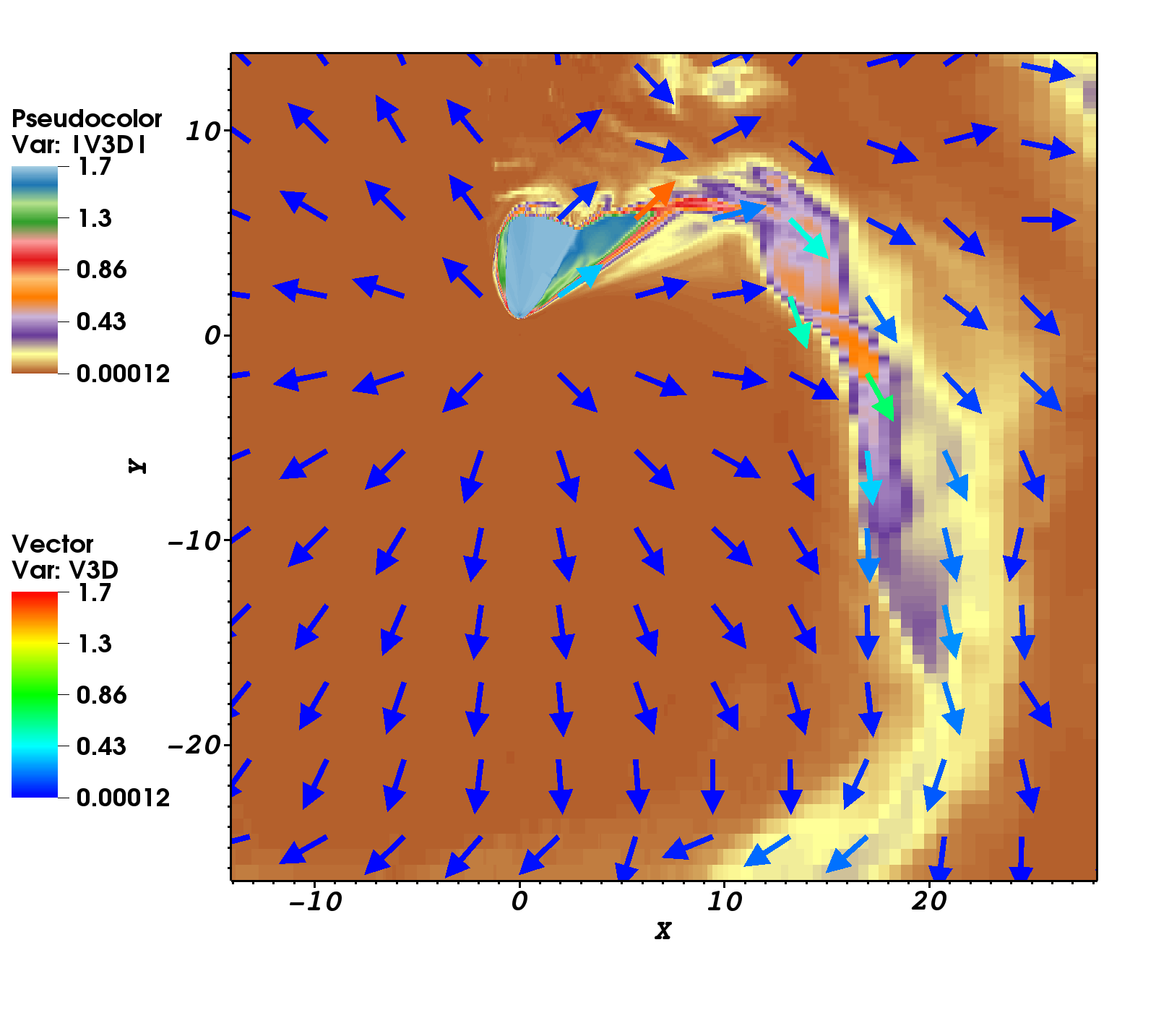}
\includegraphics[width=70mm]{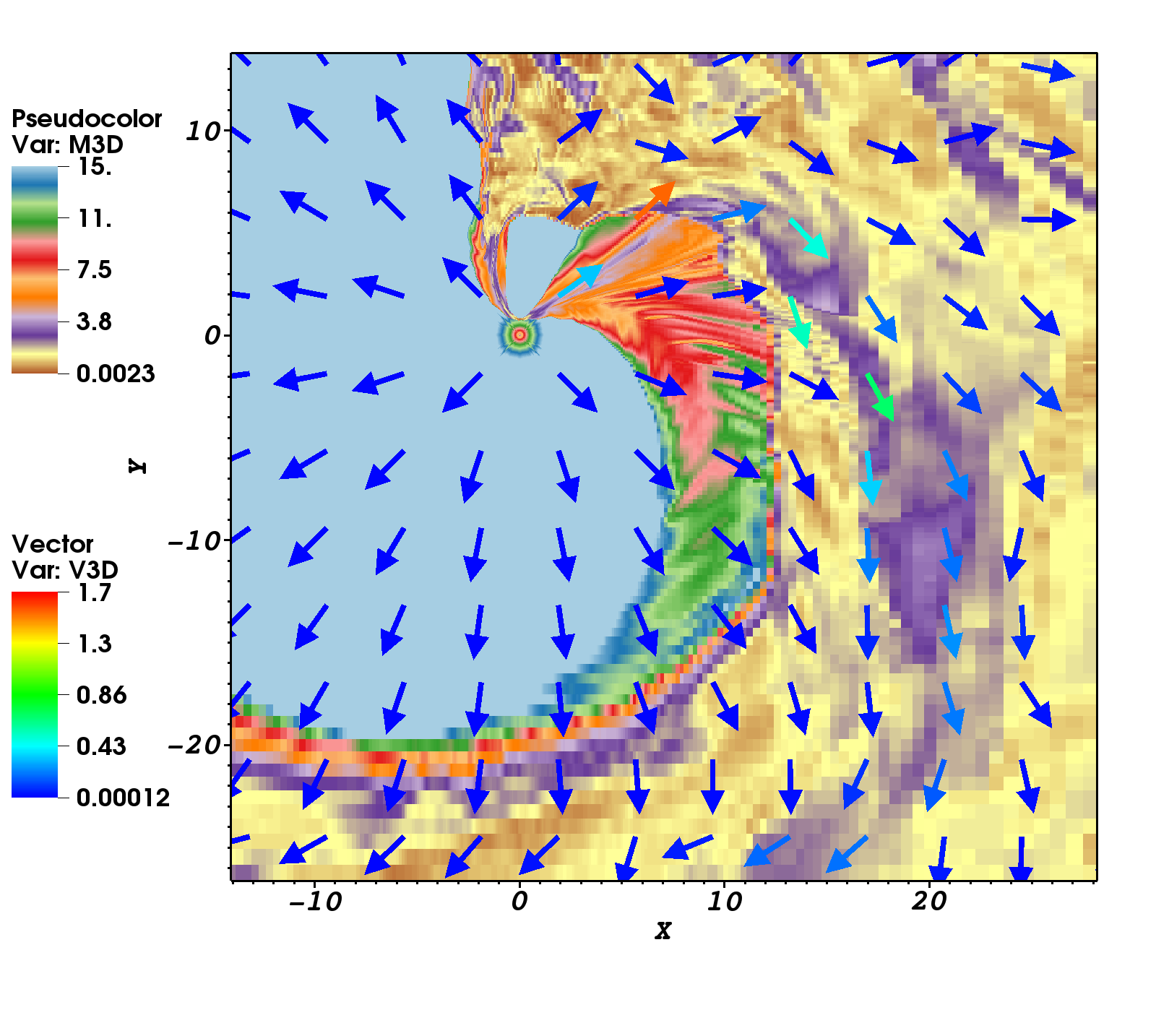}
\caption{Distribution of tracer, $\gamma\beta$, and Mach number (from top to bottom) for 3Dlf at $t=5.2$~days.}
\label{fig:3dlftum}
\end{figure}
%fffffffffffffffffffffffffffffffffffffffffffffffffffffffffffff

\subsection{Simulated physics}

The pulsar-to-star wind momentum rate ratio, $\eta=\dot{P}_{\rm pw}/\dot{P}_{\rm sw}$, was fixed to 0.1 in 3D. This value for $\eta$ is a standard value for the case of a massive star and a powerful pulsar \citep[e.g.][]{bb11,bbkp12,Dubus2013}.
For a meaningful geometric comparison between the 3D and 2D results on scales $\sim a$, we set $\eta_{\rm 2D}=\eta_{\rm 3D}^{1/2}$ to locate the two-wind contact discontinuity (CD) at the same distance from the pulsar ($R_{\rm p}$), as 
$$R_{\rm p3D}=\eta_{\rm 3D}^{1/2}\,d/(1+\eta_{\rm 3D}^{1/2})\,\,{\rm vs}\,\,R_{\rm p2D}=\eta_{\rm 2D}\,d/(1+\eta_{\rm 2D})$$ 
where $d$ is the separation distance between the two stars. Under negligible pressure, the momentum rates are 
$$\dot{P}_{\rm pw}\approx 4\,\pi\,R_{\rm p}^2\,\Gamma^2\beta^2\rho_{\rm pw}\,c^2\,\,{\rm and}\,\, \dot{P}_{\rm sw}\approx 4\,\pi\,(d-R_{\rm p})^2\rho_{\rm sw}\,v_{\rm sw}^2\,,$$ 
for the pulsar and the stellar winds, respectively; $\Gamma$, $\rho_{\rm pw}$ and $\beta$ are the pulsar wind Lorentz factor, rest-frame density, and $c$-normalized velocity; and $\rho_{\rm sw}$ and $v_{\rm sw}$ the stellar wind density and velocity. The pulsar spin-down luminosity can be approximated as 
$$L_{\rm sd}\approx (\Gamma-1)\dot{P}_{\rm pw}\,c/\beta\,.$$ Both winds are injected with spherical symmetry, and their pressure at injection can be considered negligible. In particular for the case of the pulsar wind, the pressure was chosen as low as allowed by the numerical scheme.

%fffffffffffffffffffffffffffffffffffffffffffffffffffffffffffffffffff
\begin{figure}[htp]
\centering
\includegraphics[width=94mm]{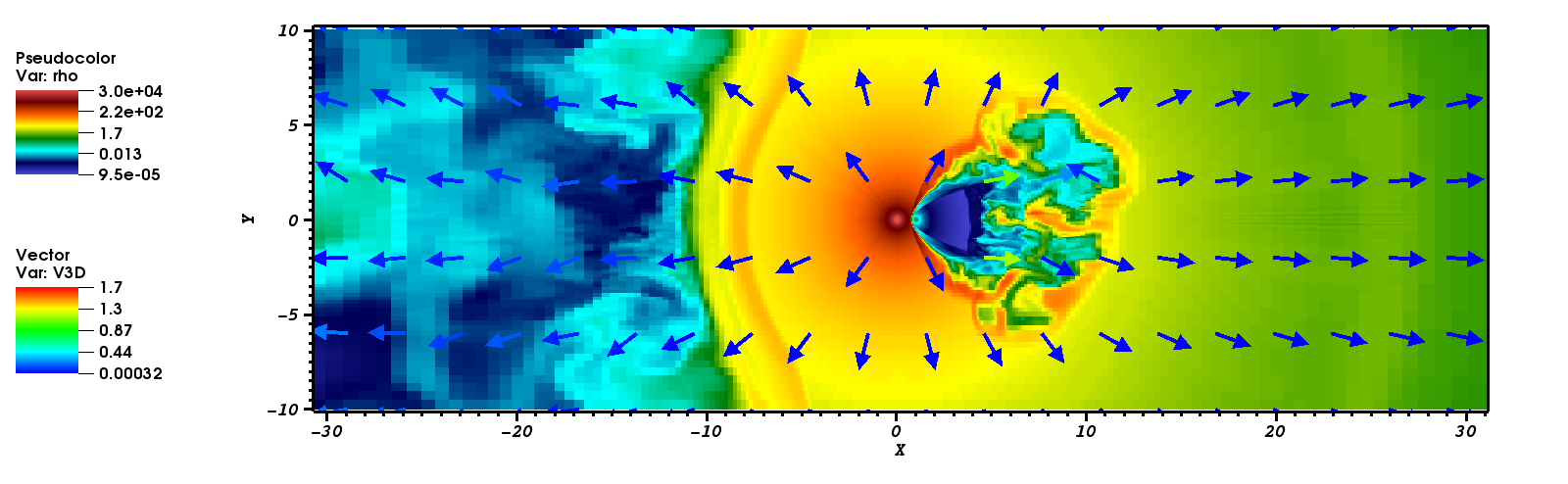}
\includegraphics[width=94mm]{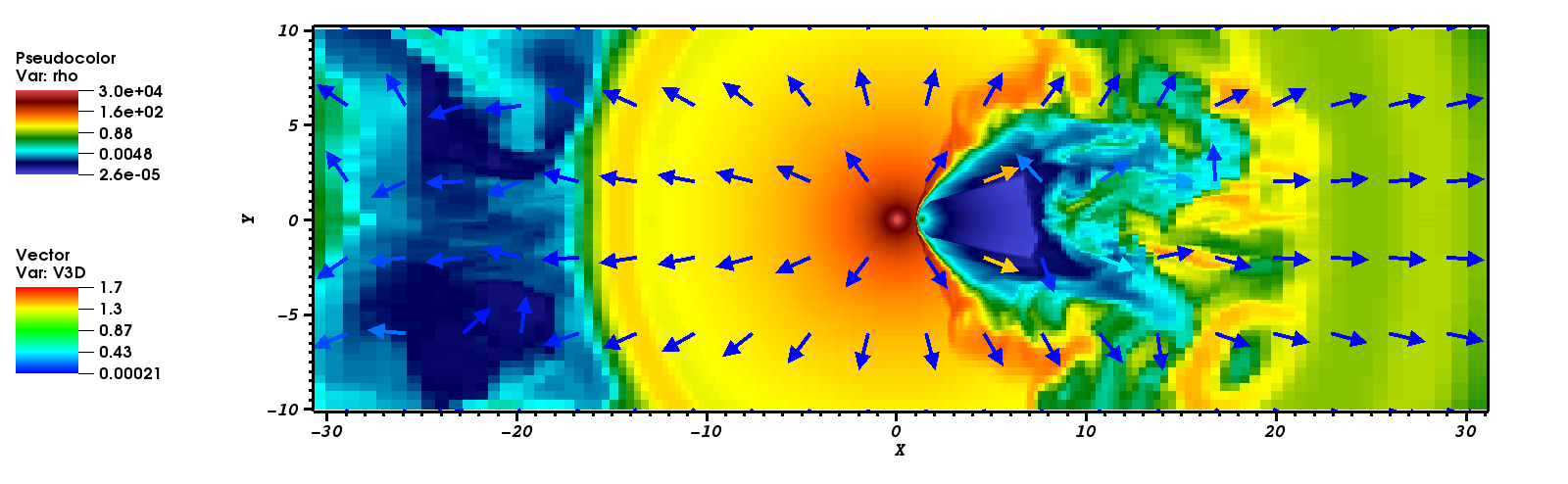}
\includegraphics[width=94mm]{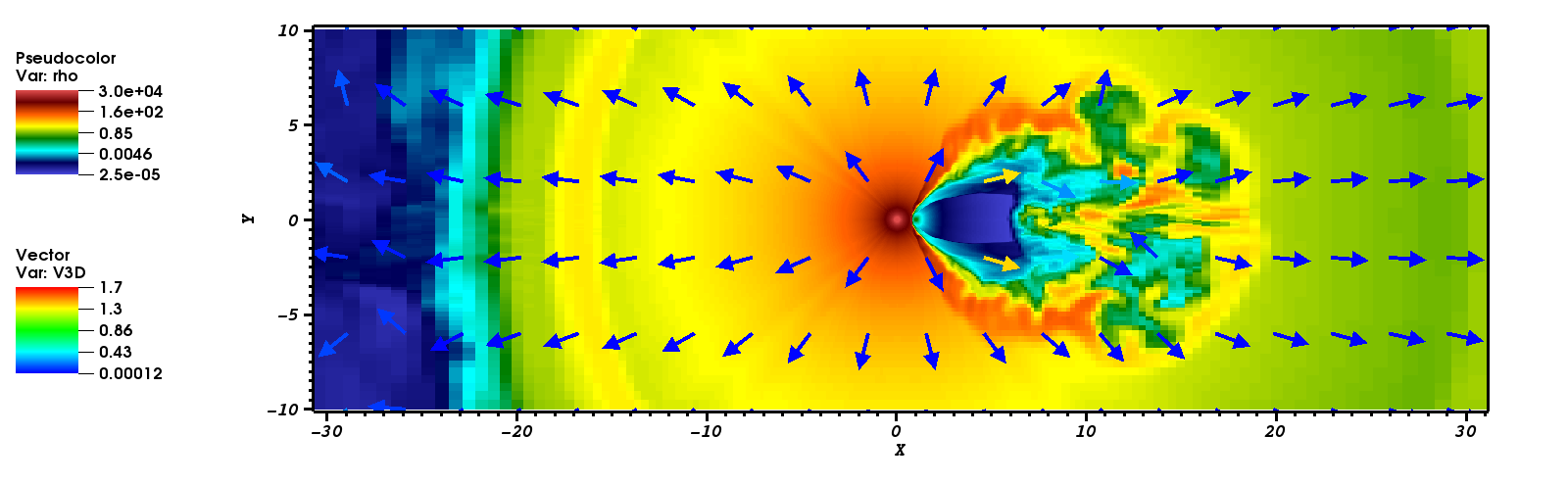}
\includegraphics[width=94mm]{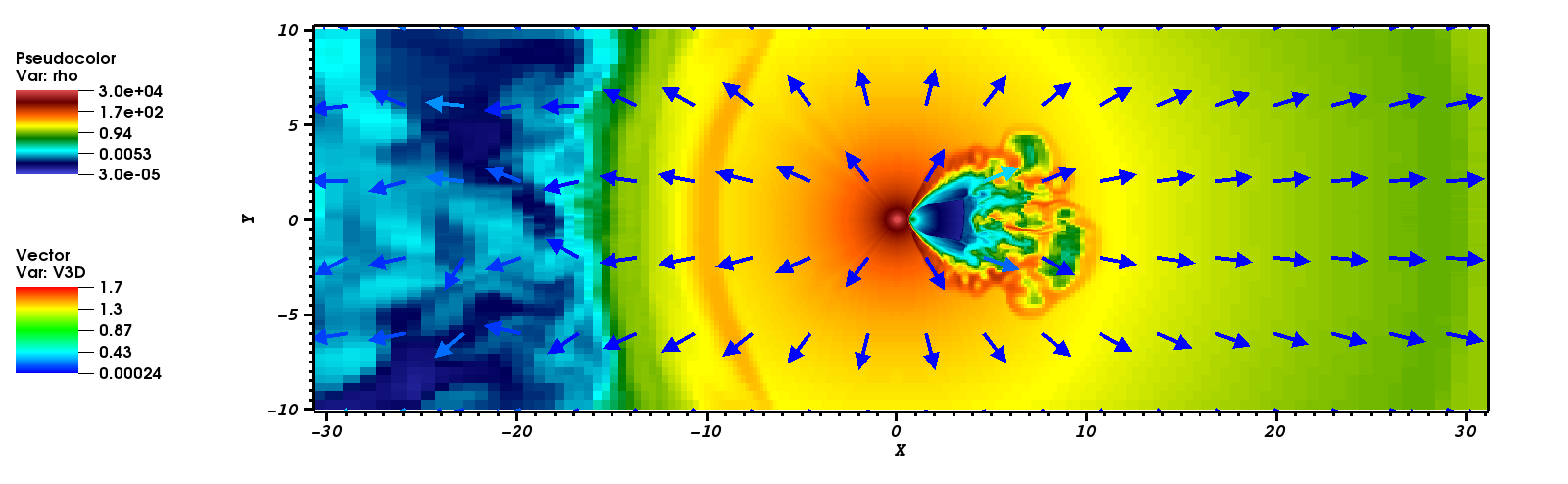}
\caption{Density distribution by colour, and arrows representing the flow motion direction in the plane perpendicular to the orbit and crossing the star-pulsar axis, for 3Dlf at:  $t=2.6$, 3.9 (apastron), 5.2, and 5.85~days (periastron) (from top to bottom).}
\label{fig:3dlfxz}
\end{figure}
%fffffffffffffffffffffffffffffffffffffffffffffffffffffffffffffffffff

A $\Gamma$-value of two was adopted. Our value of the Lorentz factor is the same as the one taken by \cite{bbkp12}, 
and it yields high wind-density and velocity contrasts in the two-wind contact discontinuity (CD), although still much higher than those implied 
by the conventional value $\Gamma\sim 10^4-10^6$ \citep[see][and references therein]{kha12,abk12}. A modest $\Gamma$-value has been chosen 
because of resolution limitations, since the reconstruction of the fluxes can lead to negative values for the term $(1-\beta^2)$ if $\beta$ is 
very close to 1. Pressure may also be affected. However, given that for $\Gamma=2$, the amount of kinetic energy is equal to the rest-mass energy of the pulsar wind, 
relativistic effects start to become apparent. On the other hand, because pulsar wind $\Gamma$-values are thought to be much higher than simulated here, 
the development of instabilities could be different. For instance, the Kelvin-Helmholtz instability (KHI) growth is proportional to the square root 
of the density contrast \citep{cha61}, and $\Gamma\sim 10^4-10^6$ would imply significantly slower growth rates. The analysis of instability growth in the present 
scenario is numerically and physically complex, so specific thorough studies beyond the scope of this work are needed, but it is worth noting that
2D simulations with $\Gamma=10$ gave qualitatively similar results to those with $\Gamma=2$ \citep[see Sect.~\ref{3dc}, and][]{bbkp12}. 
A detailed discussion of the impact of relativistic effects on the shocked-wind structure can be found in \cite{bbkp12}. In the same paper, there is also a discussion 
of the KHI development with different flow velocities \citep[see also][]{Lamberts2013}. In addition to the KHI, we also note below the development 
of the Richtmyer-Meshkov \citep[e.g.][]{richt60,m69,b02,nwm10,i12,mm13} and the Rayleigh-Taylor instabilities \citep[e.g.][]{t50,cha61}, acting together (RMI+RTI) at the orbit-leading side of the two-wind CD. The RMI+RTI, not considered before, are less dependent on the density contrast than the KHI and should not be strongly affected even if $\Gamma$ were much larger than adopted. It is worth noting that for the shocked relativistic pulsar wind, it is inertia rather than density what matters, but the stellar wind still dominates the growth of the RMI+RTI, and so it occurs in the non-relativistic regime.

Unlike \cite{bbkp12}, who simulated a generic case, in this work the orbital elements and stellar wind velocity ($v_{\rm w}$) were taken as similar to those of LS~5039 \citep{Casares2005,Aragona2009,Sarty2011}: eccentricity $e=0.24$; period $T=3.9$~days; orbital semi-major axis $a=2.3\times 10^{12}\,{\rm cm}$; and $v_{\rm w}=0.008\,c=2.4\times 10^8$, with a stellar wind Mach number of 7 at injection.
For the given $\eta$, $\Gamma$, and $v_{\rm w}$ values, the stellar and pulsar wind momentum rates,
normalized to the stellar mass-loss rate, $\dot{M}_{-7}=(\dot{M}/10^{-7}\,M_\odot\,{\rm yr}^{-1})$, are $\dot{P}_{\rm sw}\approx
6\times10^{27}\dot{M}_{-7}$ and $\dot{P}_{\rm pw}\approx 6\times10^{26}\dot{M}_{-7}$~g~cm~s$^{-2}$ ($\approx 2\times10^{27}\dot{M}_{-7}$~g~cm~s$^{-2}$ in 2D), respectively. The corresponding spin-down luminosity in 3D is $\approx 2\times 10^{37}\dot{M}_{-7}$~erg~s$^{-1}$. The values of $v_{\rm w}$, $\Gamma$, $\dot{M}$, and $\eta$ determine the wind densities. Since $\dot{M}$ is not required in these adiabatic gas simulations, its value is left unconstrained, but the adopted normalization of $10^{-7}\,M_\odot\,{\rm yr}^{-1}$ is characteristic of O-type stellar winds. 

The simulations do not include the magnetic field or anisotropy in the pulsar wind, although their impact should be small. In particular for the magnetic field, this should be the case at least when the magnetic-energy to particle-energy density ratio is $\sigma\lesssim 1$ \citep[see][]{bog12}. On larger scales, the shocked mixed winds propagate in a medium with steep drops of density and pressure \citep{bb11}, and thus $\sigma$ should remain small until the flow is terminated on pc scales. On those scales, which are much larger than those simulated here, $\sigma$ may evolve as when the pulsar wind interacts directly with the ISM or the interior of a SNR, in which case magnetic lines accumulate in the nebula, making $\sigma$ grow \citep{ken84}.

\begin{table*}
\caption{Parameters of the models considered in this work. $N_x$, $N_y$, and $N_z$ are the resolution in the $X$-, $Y$-, and $Z$-directions, respectively. 
The first number shows the resolution in the region with negative coordinates, the second one is the resolution in the uniform central region, and the third the resolution in the region with positive coordinates. $X$, $Y$, and $Z$ denote the boundaries between these three regions in each dimension; $v_{w,8}=v_{ws}/10^8 \mbox{cm s}^{-1}$. }
\begin{center}
\label{tab:models}
\begin{tabular}{|l|rrrrrrrrr|}
\hline
Name & $N_x$ & $N_y$ & $N_z$  & X & Y & Z & $v_{w,8}$ & $\eta$ & EoS \\
\hline
\hline
3Dlf & s 128 u 256 s 128 & s 128 u 256 s 128 & s 64 u 128 s 64 &   -32,-2,2,32 &   -32,-2,2,32 &   -10,-1,1,10 & 2.40 & 0.1 & CtGa \\
%3Dls & s 128 u 256 s 128 & s 128 u 256 s 128 & s 64 u 128 s 64 &   -32,-2,2,32 &   -32,-2,2,32 &   -10,-1,1,10 & 0.72 & 0.1 & CtGa \\
\hline
2Dlf & s 128 u 256 s 128 & s 128 u 256 s 128 &  &   -32,-2,2,32 &   -32,-2,2,32 &    & 2.40 & 0.3 & CtGa \\
2Dle & s 128 u 256 s 128 & s 128 u 256 s 128 &  &   -32,-2,2,32 &   -32,-2,2,32 &    & 2.40 & 0.3 & Taub \\
2Dhf & s 256 u 512 s 256 & s 256 u 512 s 256 &  &   -32,-2,2,32 &   -32,-2,2,32 &    & 2.40 & 0.3 & CtGa \\
2Dhbf & s 512 u 512 s 512 & s 512 u 512 s 512 &  &   -100,-2,2,100 &   -100,-2,2,100 &    & 2.40 & 0.3 & CtGa \\
\hline
\end{tabular}
\end{center}
\end{table*}

\subsection{Initial setup}

The simulations start running ($t=0$) when the pulsar is at apastron $(1.24\,a,0,0)$ and to the right of the star $(0,0,0)$. The orbital motion takes place
in the $XY$-plane, and set counter-clockwise. Initially, the wind of the pulsar occupies a sphere of radius $a/2$, and a region bounded by a cone 
tangent to this sphere and oriented rightwards. The stellar wind occupies the rest of the grid. The initial setup is illustrated in Fig.~\ref{setup}.
The injection region of the pulsar wind, which is injected continuously, is taken as a sphere of radius $0.07\,a$. The injection region of the stellar wind has a radius of $0.15\,a$. Both injectors are smaller than their distances to the shocked winds at any point of the orbit.
After starting the simulation, the pulsar wind injector is relocated to account for orbital motion, which is done slowly enough to avoid numerical artifacts.

The quasi-steady state of the simulation can be considered to be reached when the stellar wind shocked at $t=0$ leaves the computational domain, which occurs after a simulation running time $\sim 1/6-1/8\,\times T\sim 1/2$~days.

\section{Results}

%fffffffffffffffffffffffffffffffffffffffffffffffffffffffffffffffffff
\begin{figure}[htp]
\centering
\includegraphics[width=100mm]{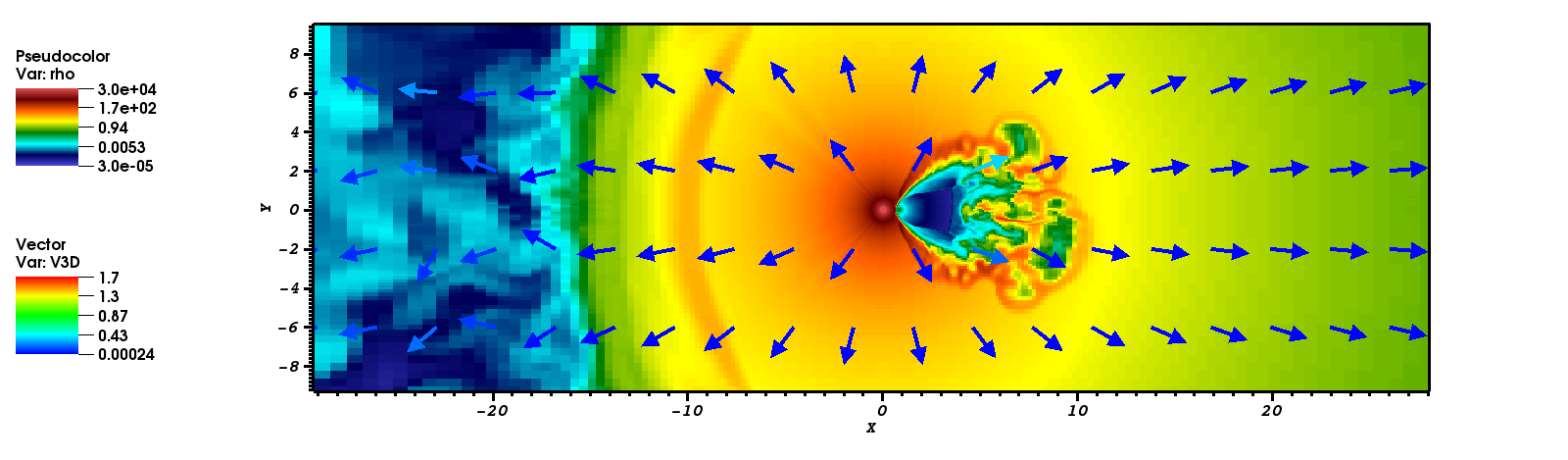}
\includegraphics[width=100mm]{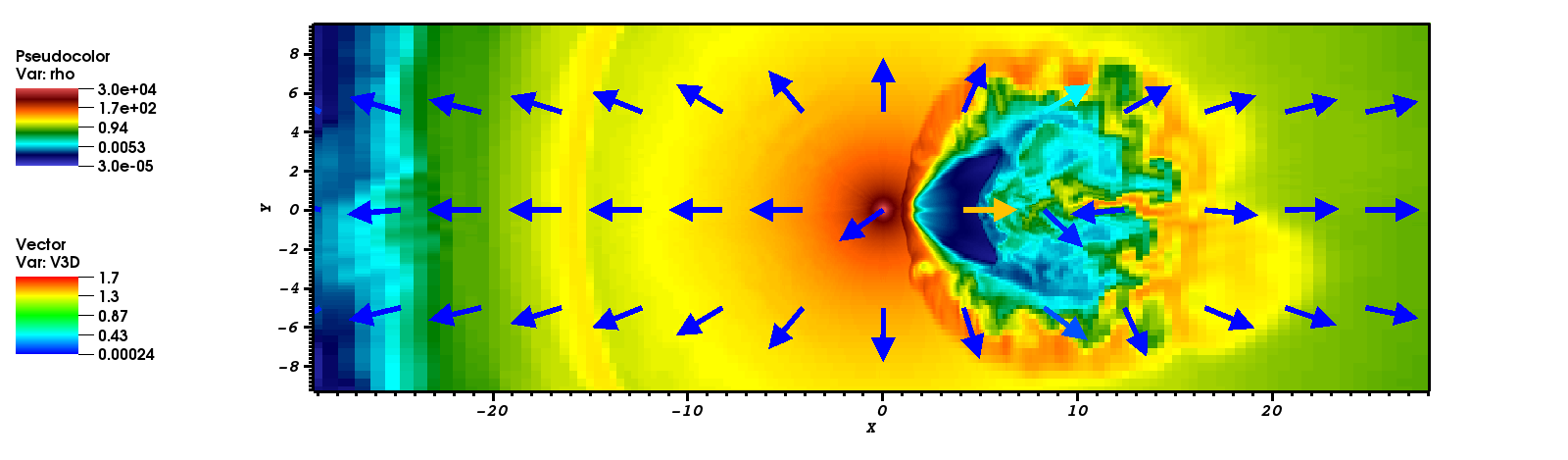}
\includegraphics[width=100mm]{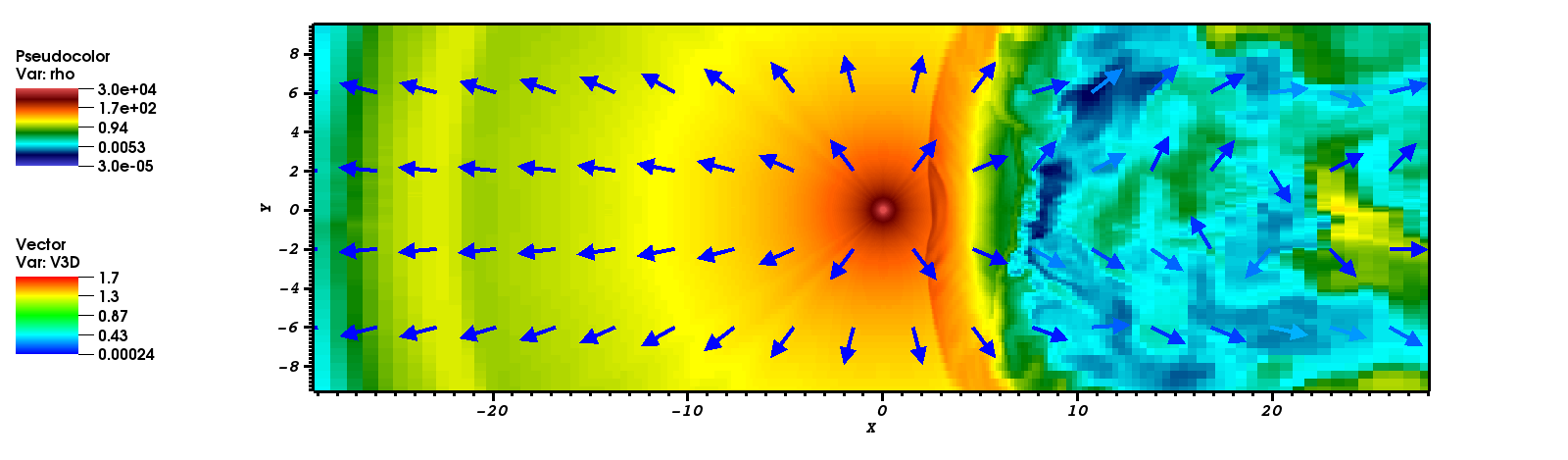}
\includegraphics[width=100mm]{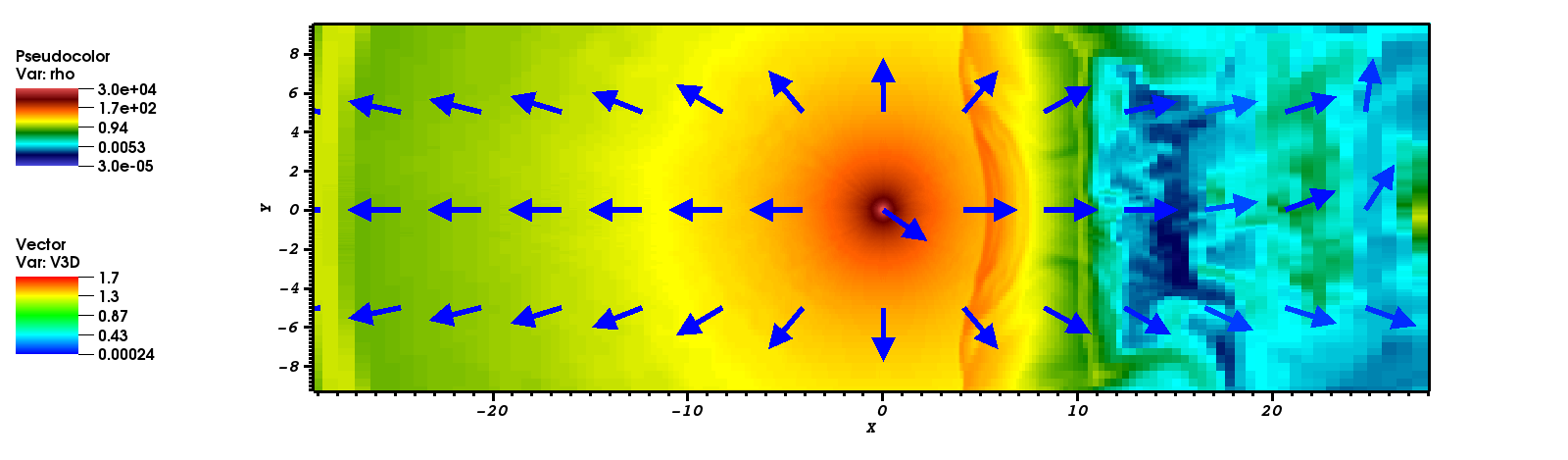}
\caption{Density distribution by colour and arrows representing the flow motion direction in a plane perpendicular to the orbit for 3Dlf at $t=5.85$~days (periastron). 
The panels show different slices with an angle $\phi$ from the star-pulsar axis and clockwise: $\phi=0,\,\pi/4,\,\pi/2,\,3\pi/4$ (from top to bottom).}
% ti3 model 
\label{fig:3dlfxzcph}
\end{figure}
%fffffffffffffffffffffffffffffffffffffffffffffffffffffffffffffffffff

%fffffffffffffffffffffffffffffffffffffffffffffffffffffffffffffffffff
%\begin{figure*}[htp]
%\centering
%\includegraphics[width=80mm]{figures/3dlf_rho_v_t246.png}
%\includegraphics[width=80mm]{figures/3dls_rho_v_t246.png}
%\caption{Comparison of the density distribution in the orbital plane (XY) for 3Dlf at $t=5.1$~days of the cases with a fast (left) and slow (right) stellar wind.  The arrows represent the %flow motion direction.}
%% ti3 model 
%\label{3Dwcomp}
%\end{figure*}
%fffffffffffffffffffffffffffffffffffffffffffffffffffffffffffffffffff

Figure~\ref{fig:3dsl} illustrates the 3D distribution of the density at apastron. The figure shows plane cuts in the $XY$-, $XZ$-, and $YZ$-planes 
taken after one full orbital period at $t=3.9$~days. The shocked-wind structure has already reached a quasi-stationary state. Figure~\ref{fig:3Dlfxyoph} shows the distribution of density in the orbital plane for 3Dlf at times $t=2.6$, 3.9 (apastron), 5.2, and 5.85~days (periastron). For better visualizing of the results of 3Dlf, a zoom in of the density distribution around the pulsar during periastron is shown in Fig.~\ref{fig:3Dzoom}. 
Also, Fig.~\ref{fig:3dlftum} shows the distribution of tracer (1 for the pulsar and -1 for the stellar wind, respectively), modulus of the four-velocity spatial component ($\gamma\beta$), and Mach number, for 3Dlf in the same plane. For completeness, Fig.~\ref{fig:3dlfxz} presents the density distribution for plane cuts along the star-pulsar axis and perpendicular to the orbital plane at the same times as those in Fig.~\ref{fig:3Dlfxyoph}. In addition, 
Figure~\ref{fig:3dlfxzcph} provides the density distribution at periastron for plane cuts perpendicular to the orbital plane at different angles from the star-pulsar axis and clockwise: $\phi=0,\,\pi/4,\,\pi/2,\,3\pi/4$. 
%For comparison of the two 3D models, Fig.~\ref{3Dwcomp} presents two snapshots at the same orbital phase ($t=5.1$~days), one corresponding to 3Dlf, and the other to 3Dls, which %has a stellar wind $\approx 3$ times lower velocity and higher density than in 3Dlf.

\begin{figure*}[htp]
\centering
\includegraphics[width=80mm]{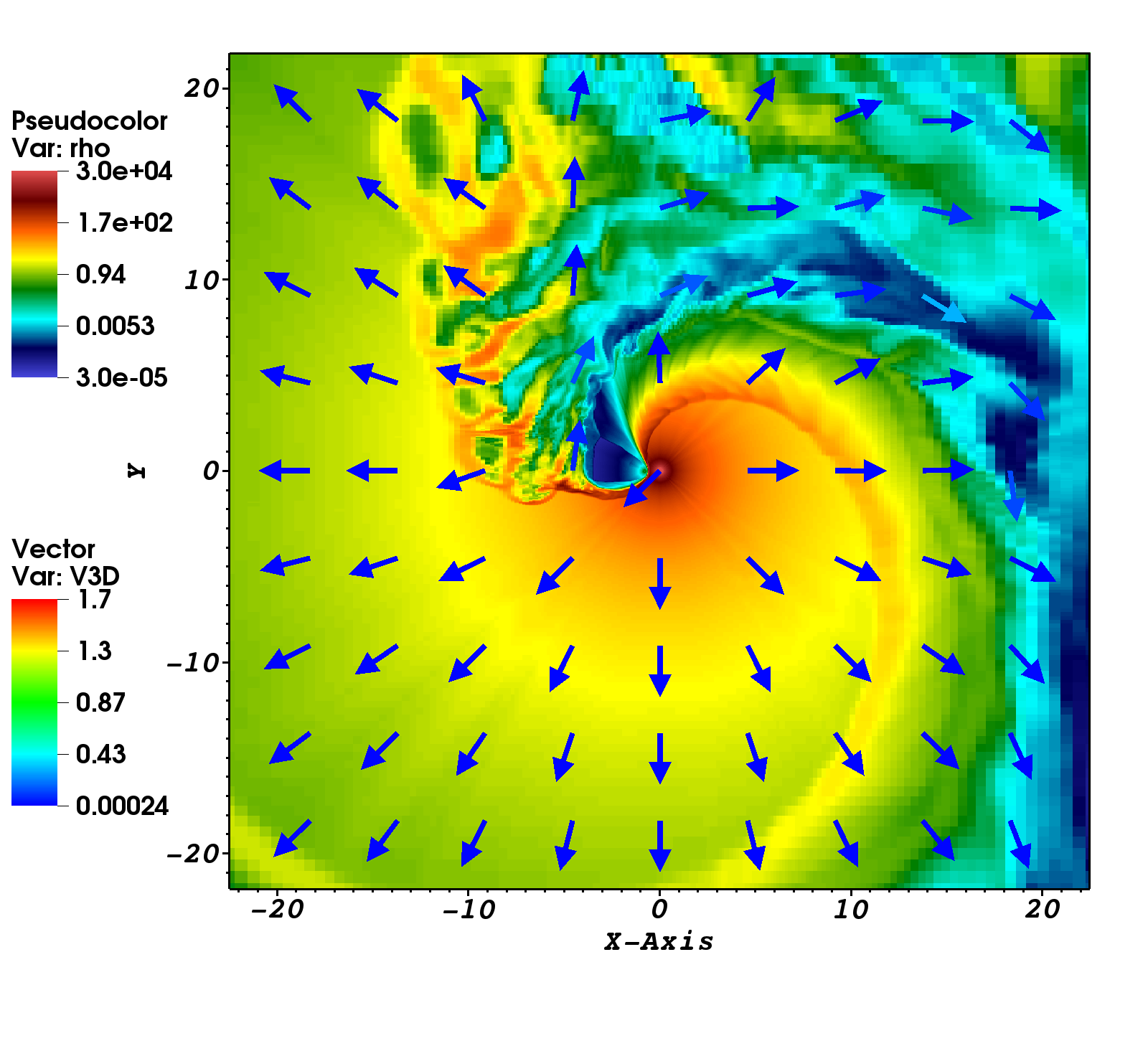}
\includegraphics[width=80mm]{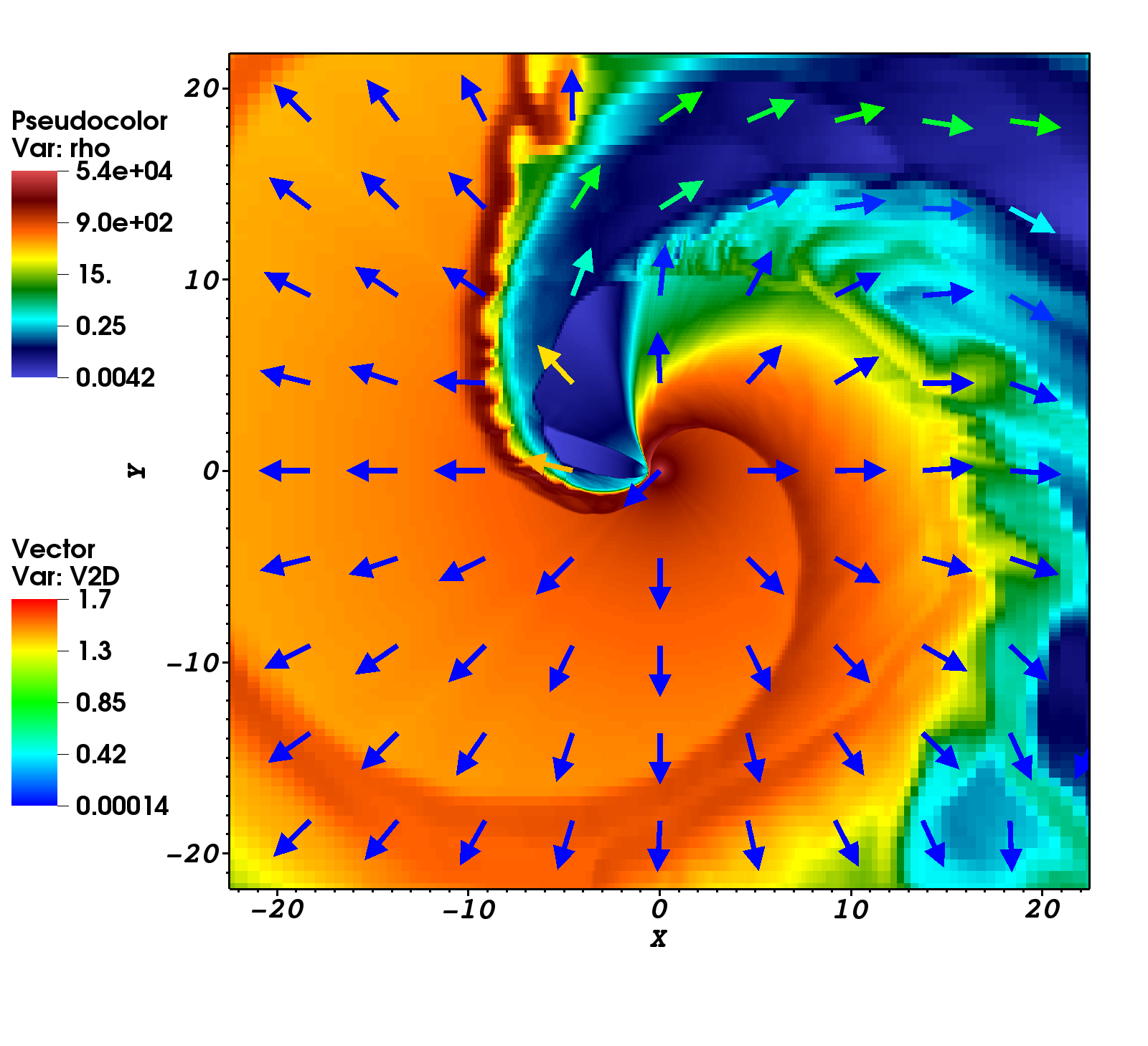}
\includegraphics[width=80mm]{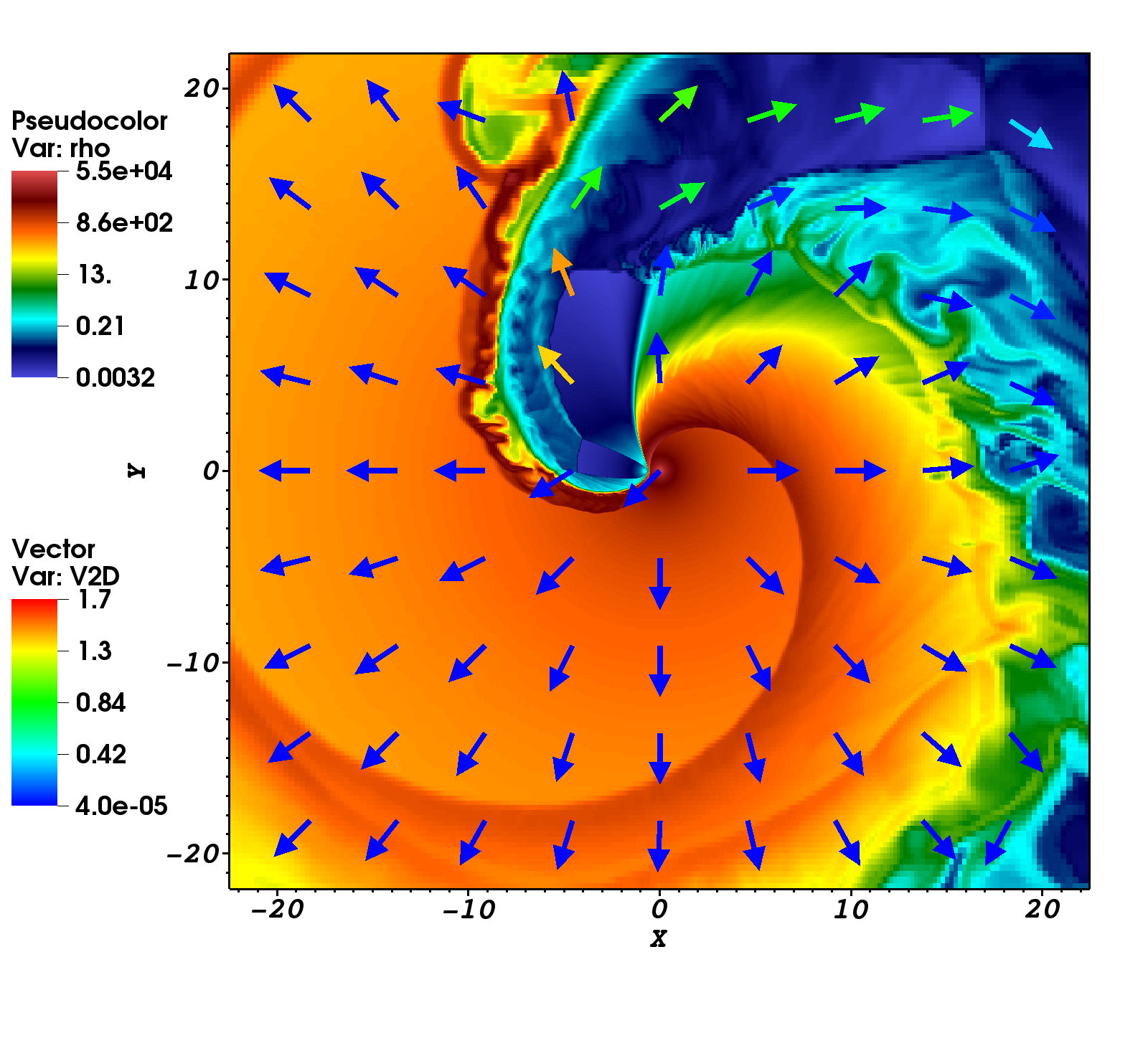}
\includegraphics[width=80mm]{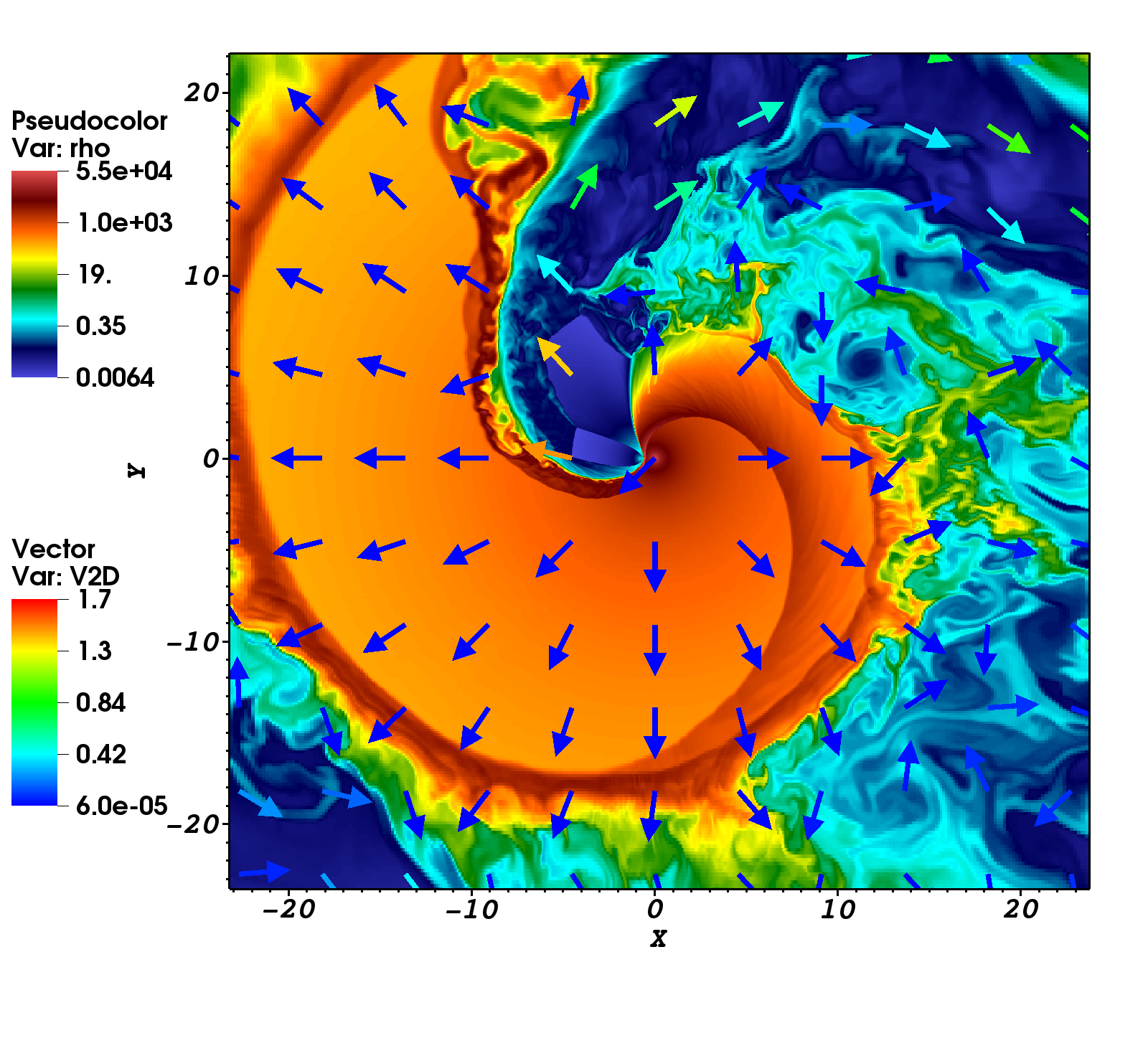}
\caption{Comparison of the density distribution (colour), and flow motion direction (arrows), in the orbital plane (XY) at $t=5.85$~days (periastron) for 3Dlf, 2Dlf, 2Dhf, and 2Dhbf 
(from left to right, and top to bottom).}
% ti3 model 
\label{fig:comparison}
\end{figure*}

\begin{figure*}[htp]
\centering
\includegraphics[width=80mm]{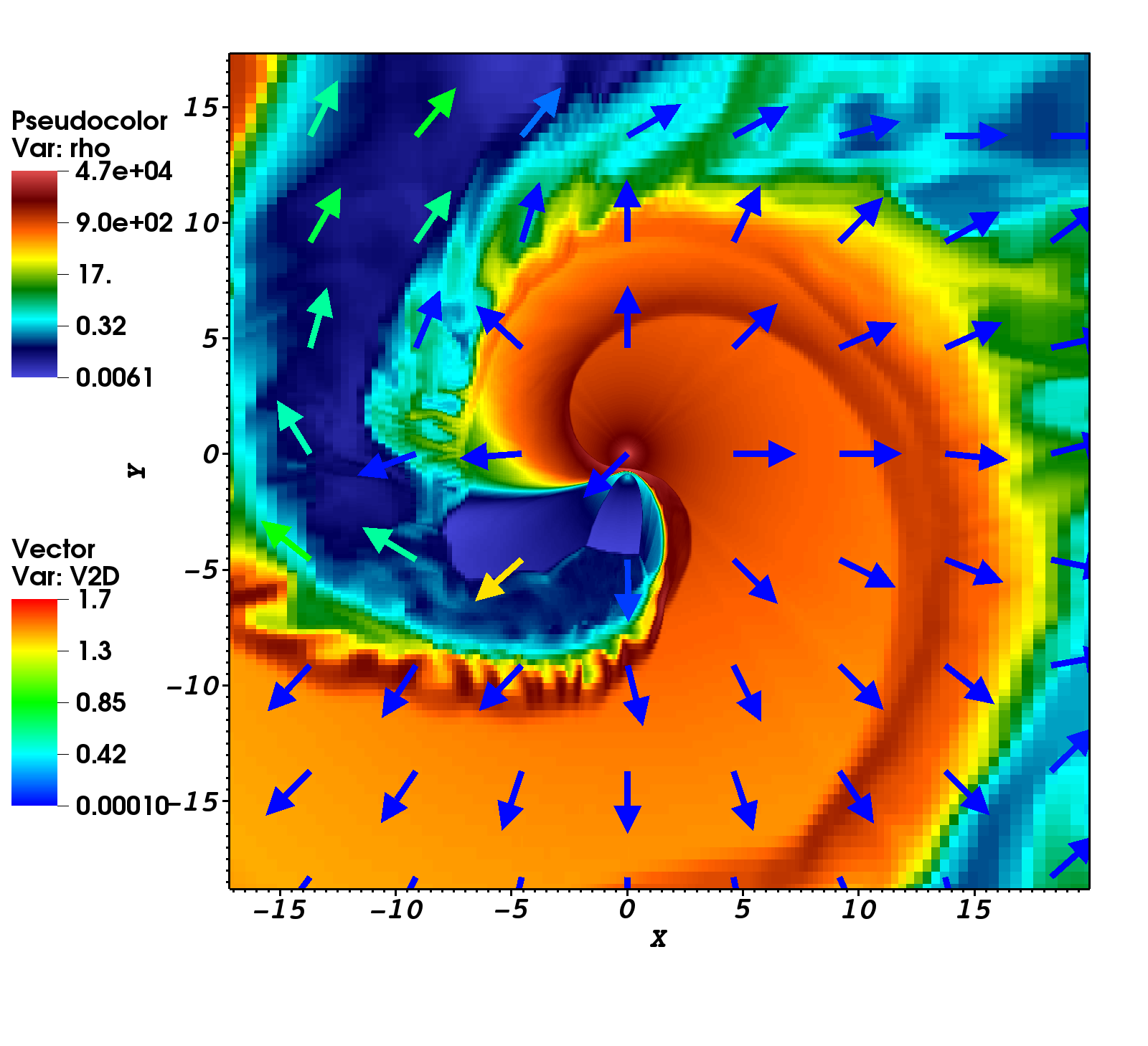}
\includegraphics[width=80mm]{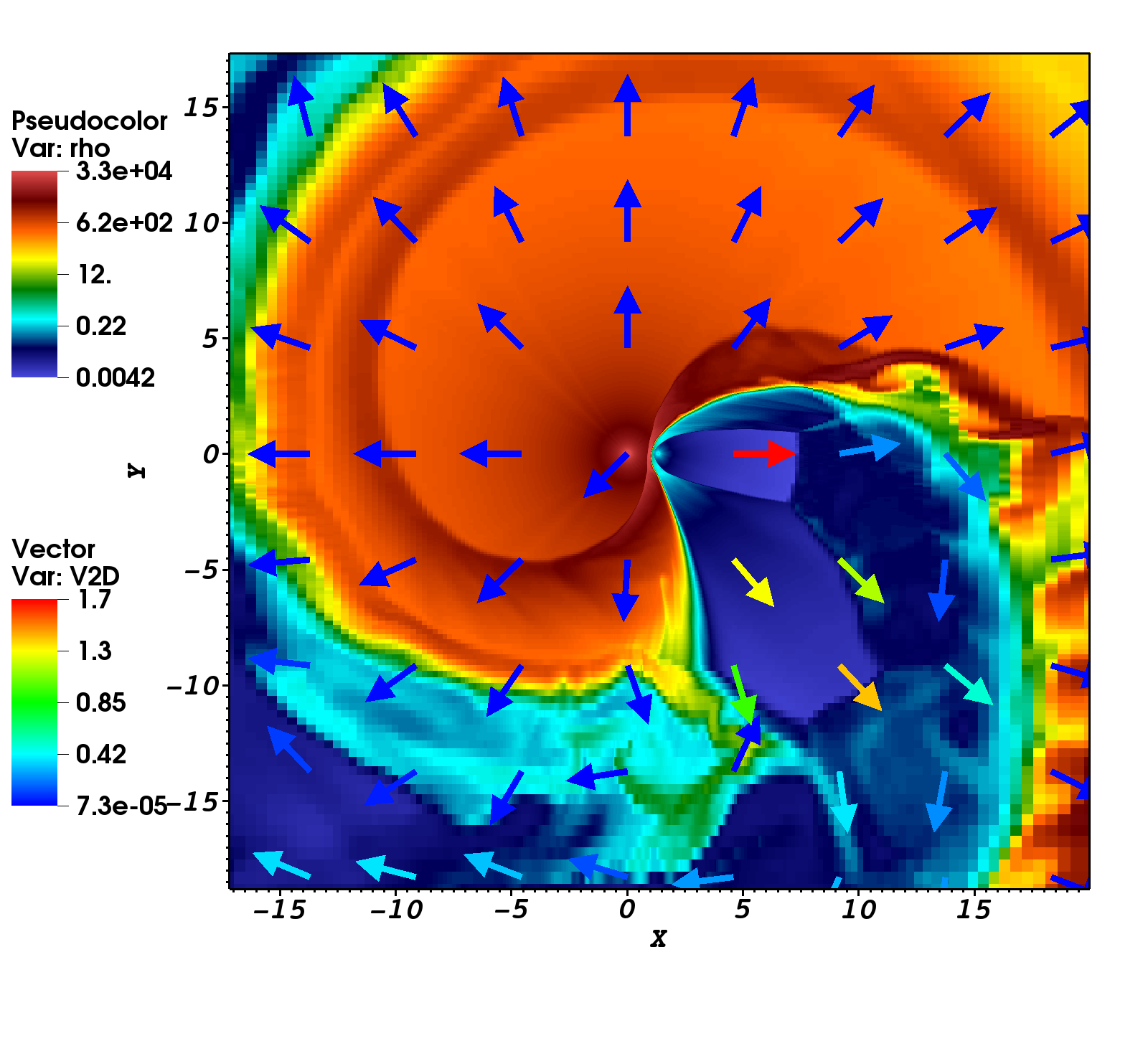}
\includegraphics[width=80mm]{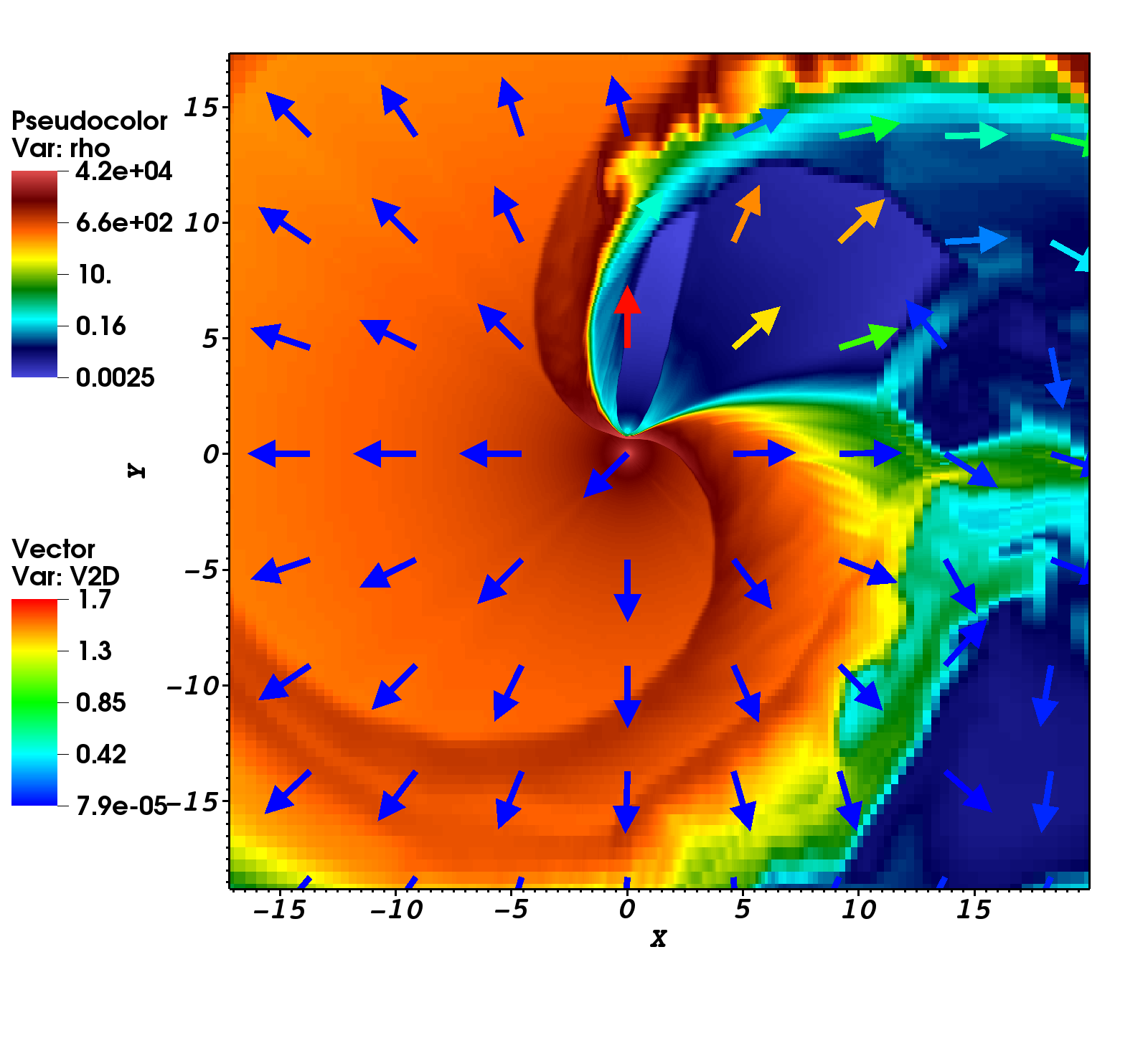}
\includegraphics[width=80mm]{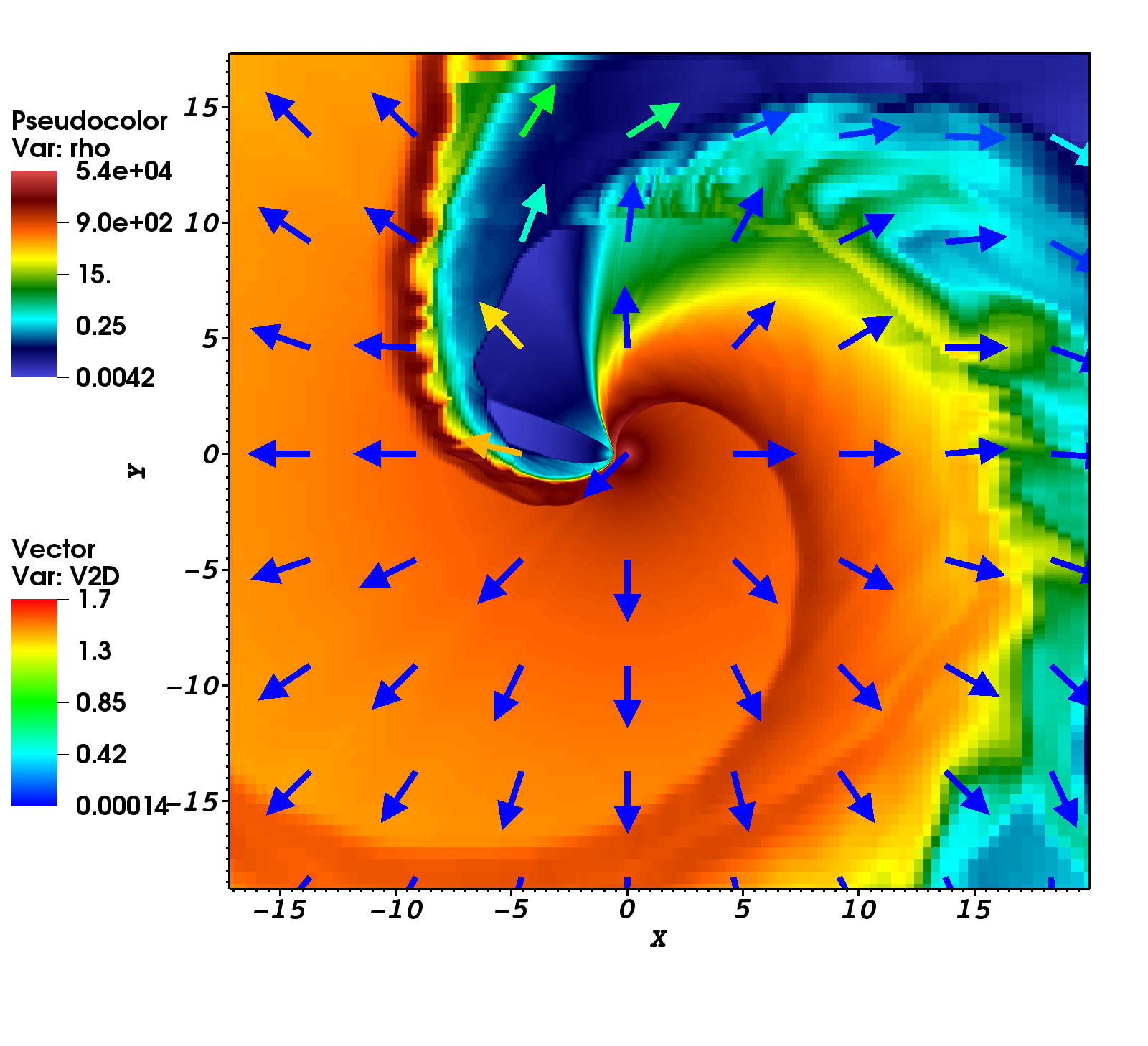}
\caption{Density distribution by colour, and arrows representing the flow motion direction, for 2Dlf at: $t=2.6$, 3.9 (apastron), 5.2, and 5.85~days (periastron) (from left to right, and top to bottom).}
% ti3 model 
\label{fig:2Dlfxyoph}
\end{figure*}

\begin{figure*}[htp]
\centering
\includegraphics[width=80mm]{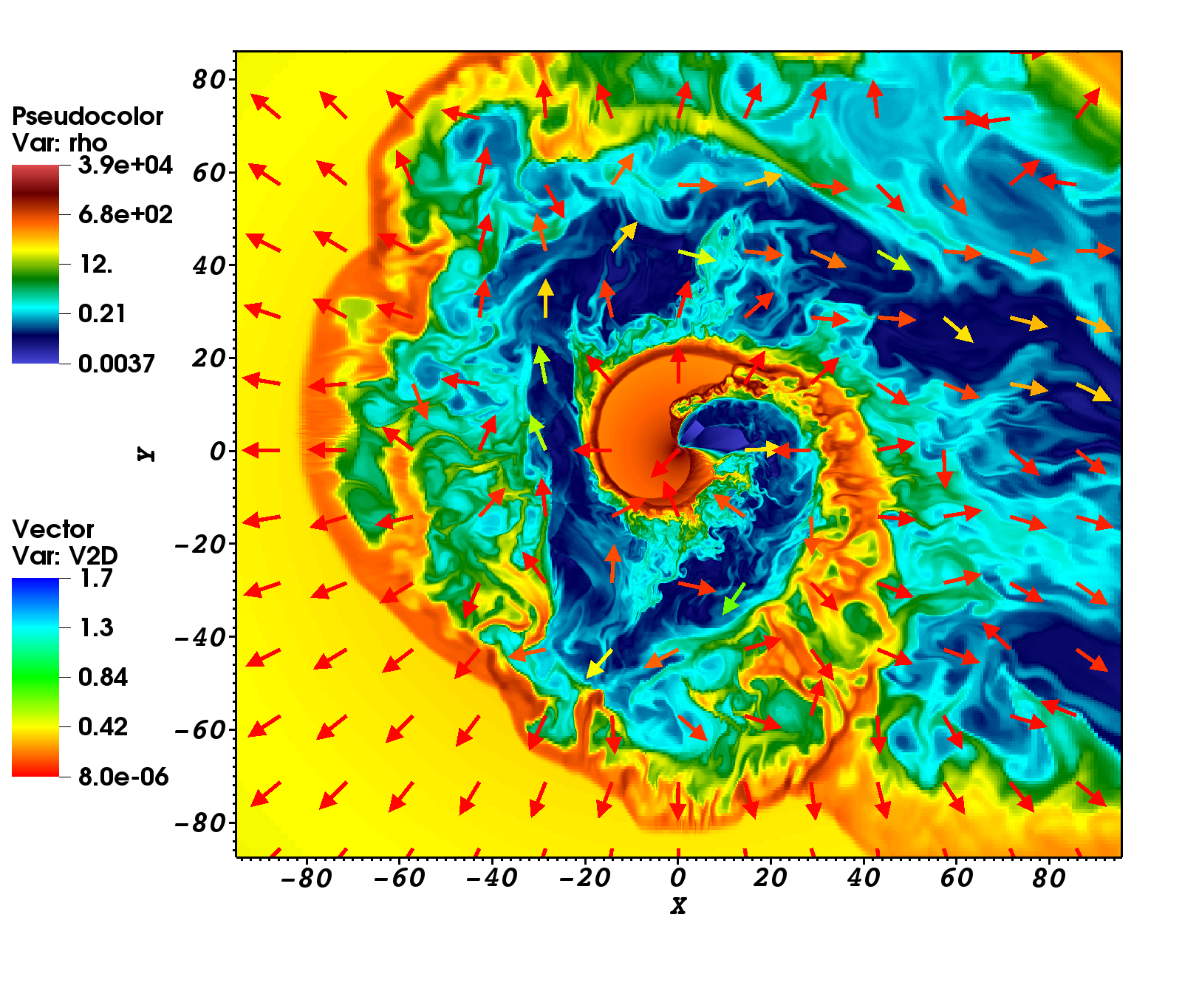}
\includegraphics[width=80mm]{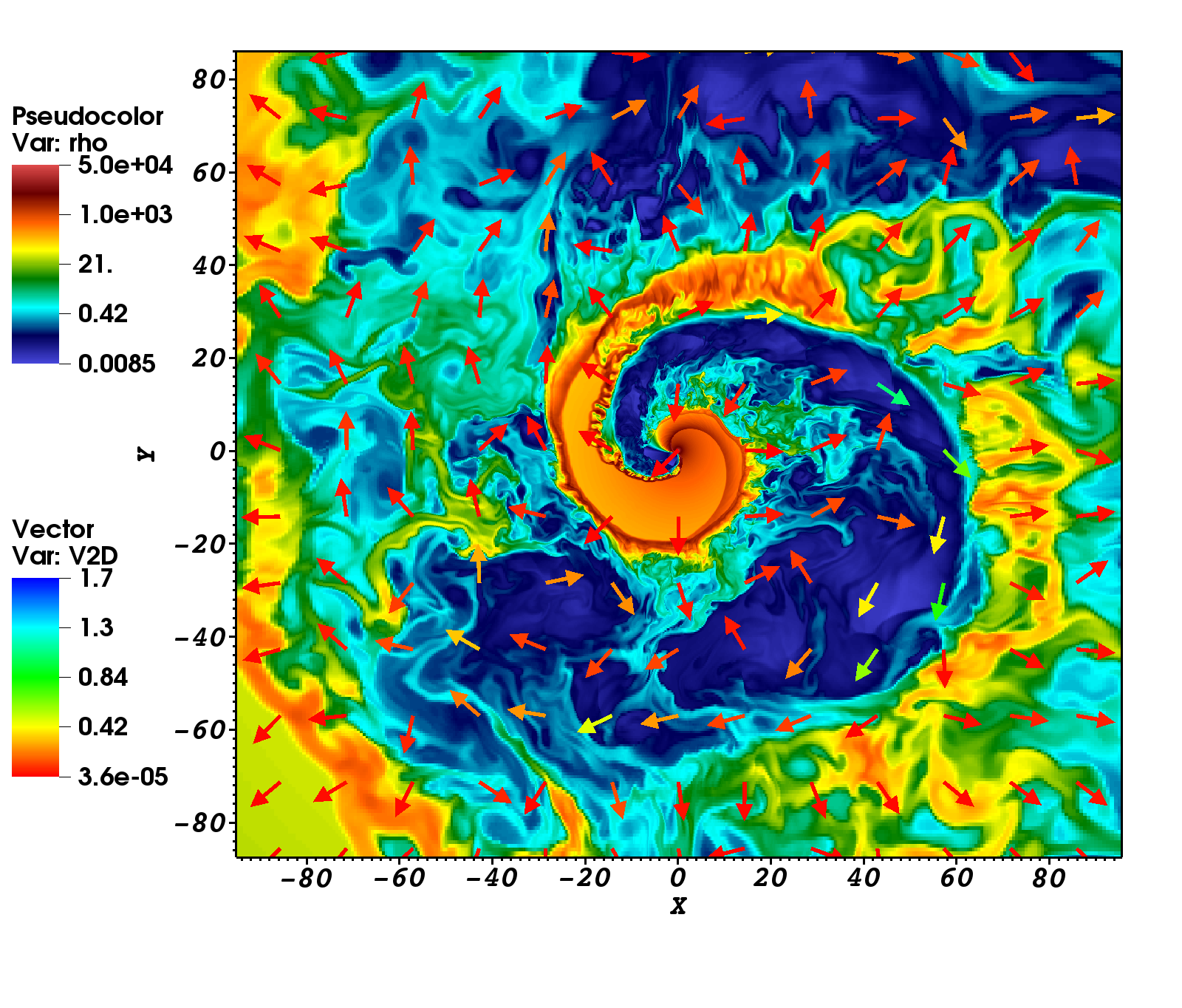}
\includegraphics[width=80mm]{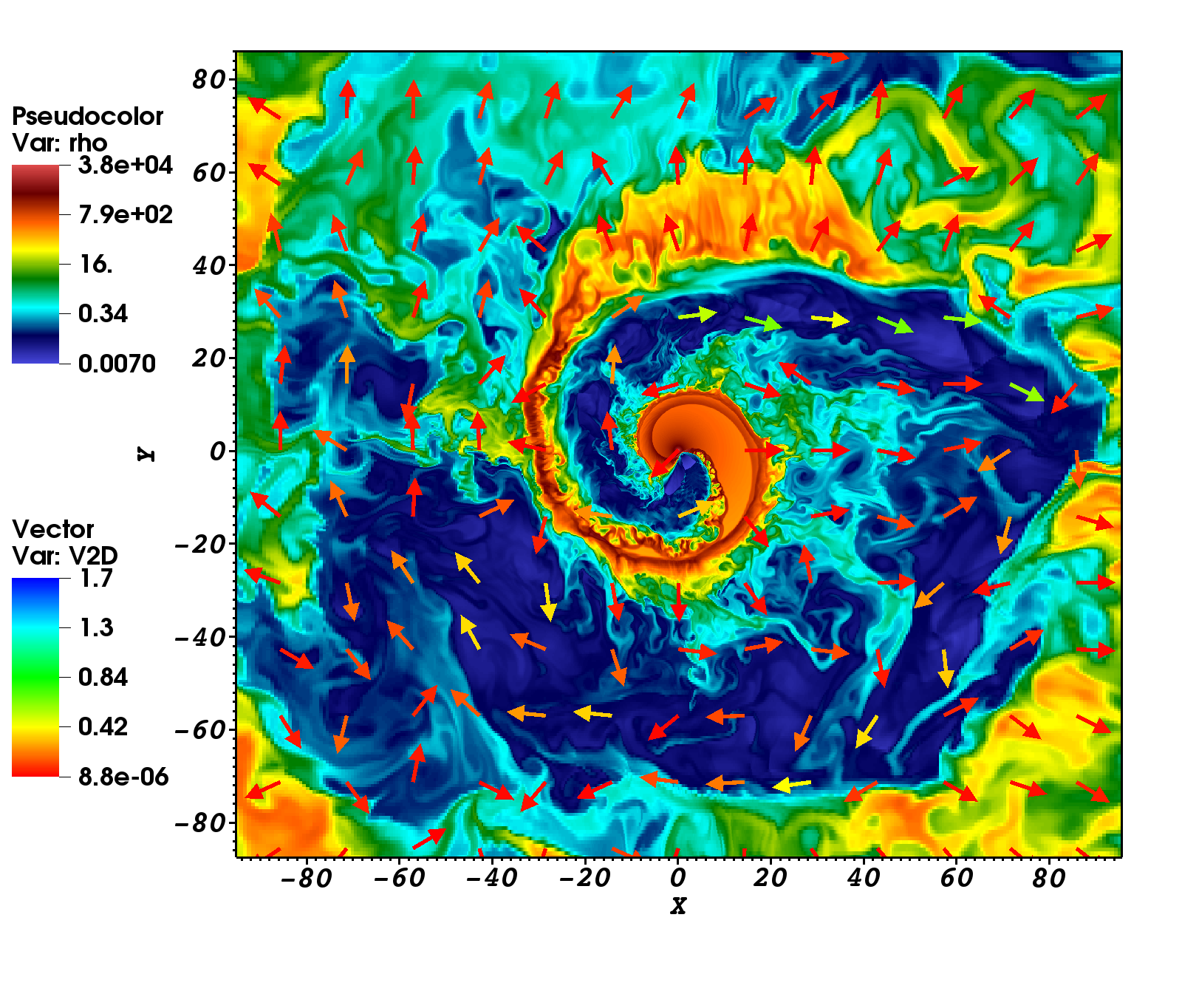}
\includegraphics[width=80mm]{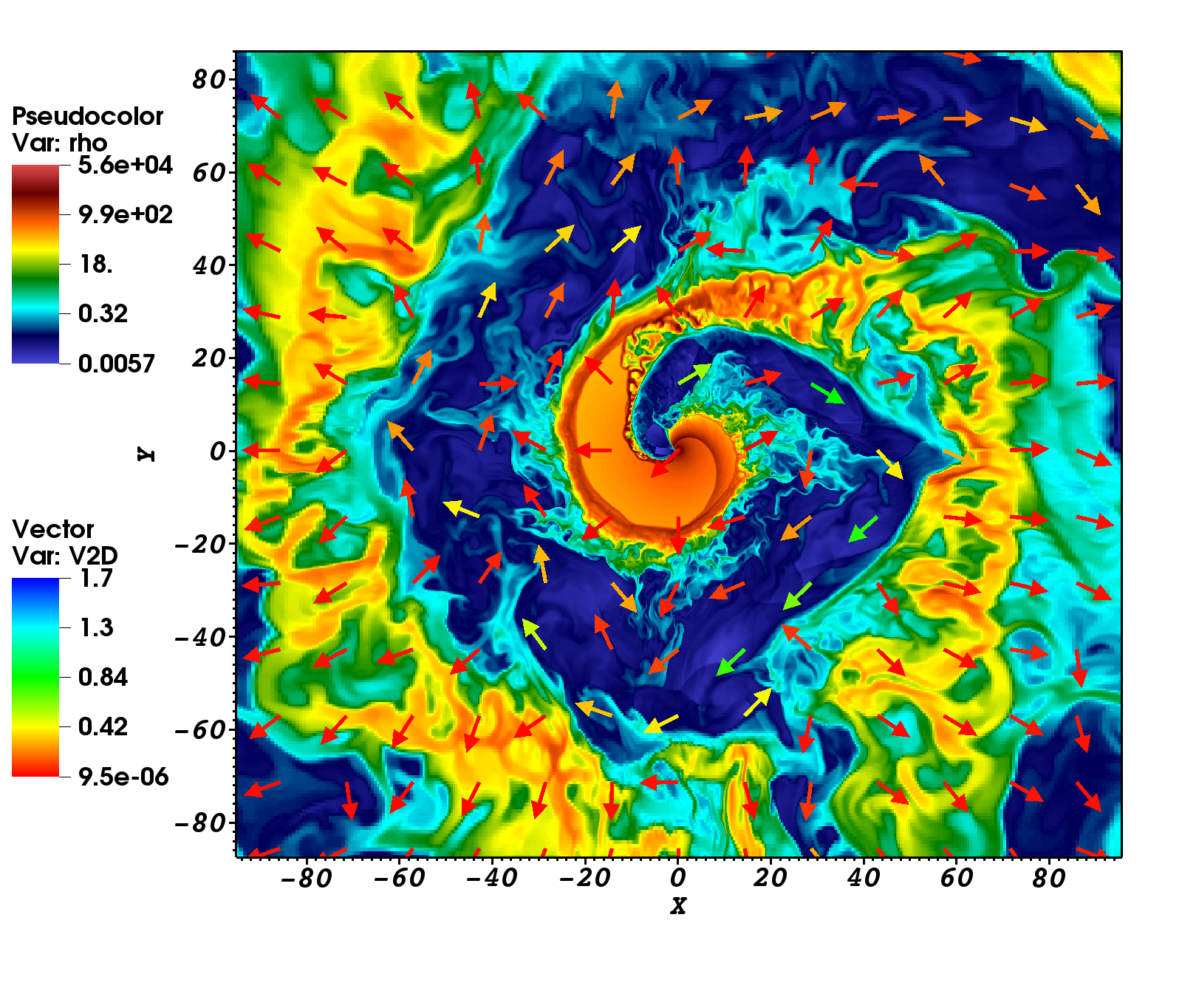}
\caption{Density distribution by colour, and arrows representing the flow motion direction, for 2Dhbf at: $t=4.8$, 6.3, 7.0, and 9.85~days ($\sim$ periastron) (from left to right, and top to bottom).}
% ti3 model 
\label{fig:2Dhbrho}
\end{figure*}

Figure~\ref{fig:2Dlfxyoph} shows the distribution of density on the orbital plane for the same times as those in Fig.~\ref{fig:3Dlfxyoph} for model 2Dlf.  The same is shown in Fig.~\ref{fig:2Dhbrho}, but for a simulation with increased resolution and grid size (2Dhbf), and wind slightly different orbital phases. Figure~\ref{fig:2dhbftum} shows the distribution of tracer, $\gamma\beta$, and Mach number, for 2Dhbf. Figure~\ref{compeos} allows for the comparison when using two different EoS: CtGa and Taub. Finally, Fig.~\ref{fig:comparison} allows the comparison of the density distribution at the same orbital phase ($t=5.85$~days) for the models 3Dlf, 2Dlf, 2Dhf, and 2Dhbf (see Table~\ref{tab:models}).

\subsection{3D case}\label{3dc}

%fffffffffffffffffffffffffffffffffffffffffffffffffffffffffffffffffff
%fffffffffffffffffffffffffffffffffffffffffffffffffffffffffffffffffff

%fffffffffffffffffffffffffffffffffffffffffffffffffffffffffffffffffff
%fffffffffffffffffffffffffffffffffffffffffffffffffffffffffffffffffff

The quasi-stationary solution of the 3D simulation confirms what has already been seen in 2D using planar coordinates in \cite{bbkp12}. Despite some quantitative differences discussed below, Figs.~\ref{fig:3dsl}, \ref{fig:3Dlfxyoph}, \ref{fig:3Dzoom}, and \ref{fig:3dlftum} display the same features as those found by \cite{bbkp12}. 

%fffffffffffffffffffffffffffffffffffffffffffffffffffffffffffffffffff
%fffffffffffffffffffffffffffffffffffffffffffffffffffffffffffffffffff

\subsubsection{Coriolis forces and instabilities}

On one hand, instabilities starting at $\sim 2\,a$ from the pulsar quickly develop in the CD. These instabilities may have two different origins. Owing to the velocity difference between the two shocked winds, their origin may be the KHI. Also, because the two media have very different densities and are subject to recurrent lateral shaking and persistent acceleration provided by the Coriolis force, the instability may be driven mostly by RMI+RTI. All these instabilities could couple together, and also with the orbital motion-induced Coriolis force, 
which pushes material from the CD that stops the propagation of the shocked pulsar wind coming from within the binary. This is apparent, for instance, at the leading edge of the shocked-wind structure at the point where the CD widens due to the instabilities (see Fig.~\ref{fig:3Dzoom}). 

The importance of RMI+RTI is also suggested by the presence of turbulence in the leading edge of the shocked flow structure, and by its almost complete absence in the traling edge. The process can be explained as follows. As the pulsar orbits the star, the underdense shocked pulsar wind penetrates the much denser shocked stellar wind, causing the RTI to develop. Because the pulsar wind shock rotates in the direction of the CD, it can catch up with the RTI fingers and trigger the development of the RMI. 
After termination within the binary, the shocked pulsar wind reaccelerates to become supersonic, and then a strong shock forms again when its 
path is blocked by the disrupted CD, heavily loaded with shocked stellar wind. The mass load of the pulsar wind region is very apparent in the figures 
as abrupt and patchy changes in density and tracer starting at $\sim 5-10\,a$ from the pulsar. This material from the CD penetrates further, stopping 
the unshocked pulsar wind moving away from the star with an almost perpendicular shock. Unlike the KHI, the RMI+RTI are less dependent on the density contrast and could have  a similar strength even for much higher $\Gamma$-values.

Despite the relevant role of the hydrodynamical instabilities, we note that the Coriolis force is the main factor that shapes the interaction region, 
inducing the strong asymmetry of the shocked-wind structure, as seen in Fig.~\ref{fig:3Dzoom}. It is the very effect of the Coriolis force that makes the 
shocked flow propagation outwards strongly non-ballistic, creating quasi-perpendicular shocks at the termination of the pulsar wind {\it behind} 
the pulsar (as seen from the star). Once shocked, the flow deviates strongly, following the trend imposed by the material against which it shocked, 
starting the spiral pattern (counter-orbitwise). These shocks do not have the same nature as the oblique shocks terminating the pulsar wind {\it behind} the pulsar 
found in axisymmetric, non-relativistic and relativistic simulations of the same scenario but without orbital motion \citep[e.g.,][]{bog08,lamberts2011}. 
In particular, in axisymmetric cases the termination shock {\it behind} the pulsar disappears for $\eta\gtrsim 0.012$, whereas here $\eta=0.1$. 
The {\it Coriolis shocks} found in our simulations are also different from those generated {\it behind} the pulsar when this is moving in the 
ISM \citep[e.g.][]{bucciantini2005}, since the environment in that case has a constant density. To finish with, as noted below, the size of the whole 
shocked-wind structure clearly depends on the orbital phase, which would be difficult to explain if the KHI alone played a dominant dynamical role. 
Because the RMI+RTI are fueled by the Coriolis force, these instabilities are actually actively participating in shaping the asymmetric shape 
of the whole two-wind interaction region.

%fffffffffffffffffffffffffffffffffffffffffffffffffffffffffffffffffff
\begin{figure}[htp]
\centering
\includegraphics[width=70mm]{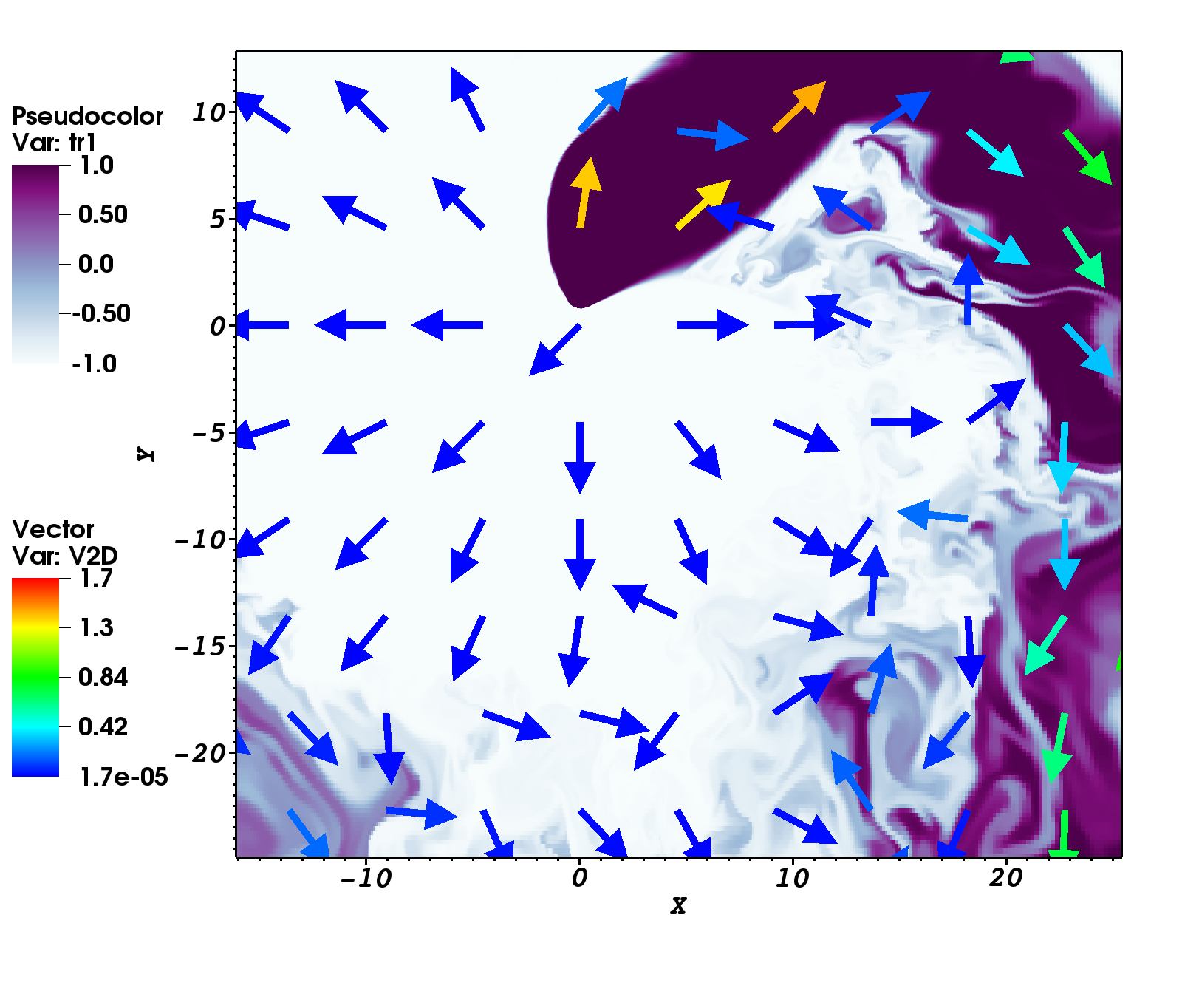}
\includegraphics[width=70mm]{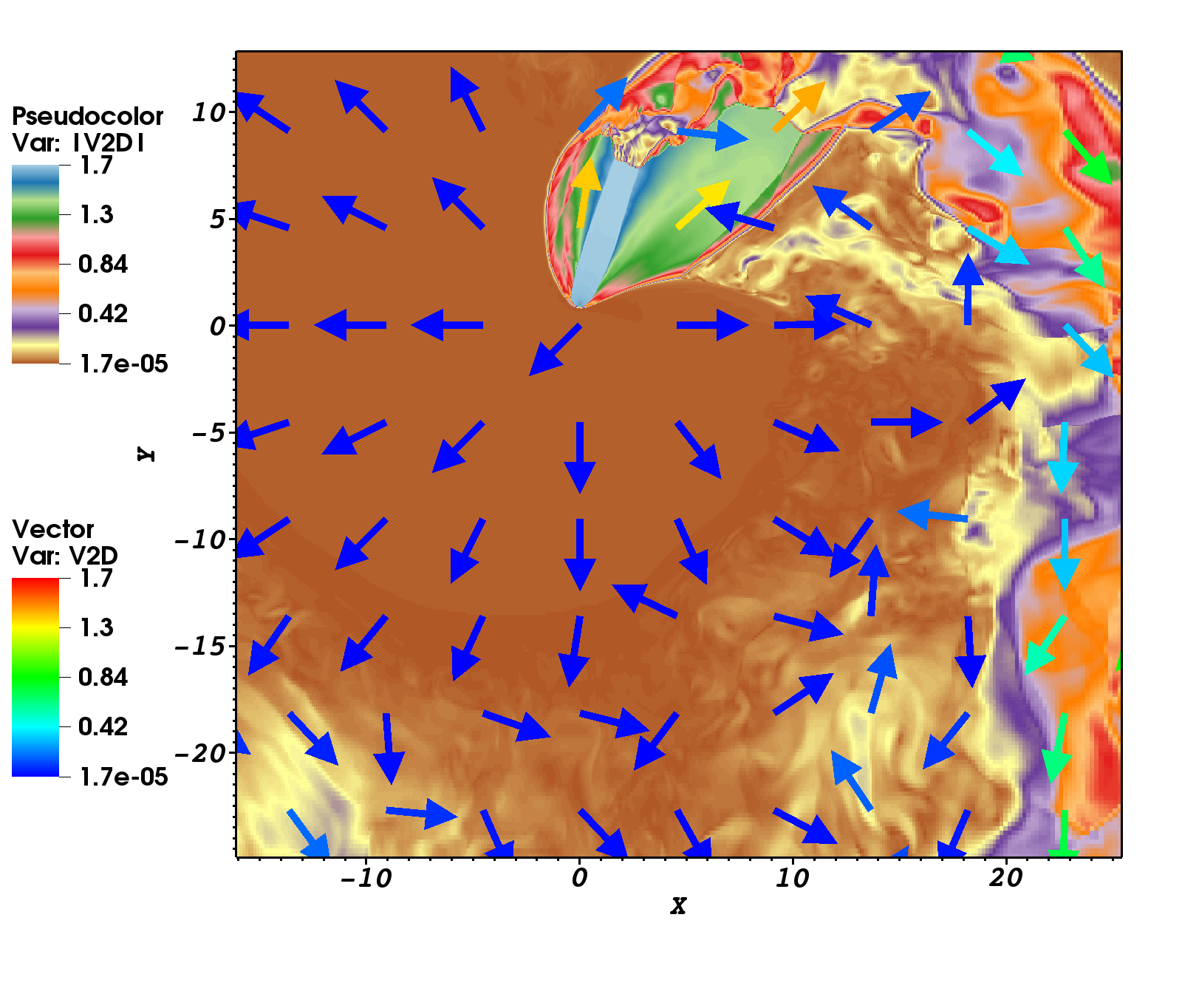}
\includegraphics[width=70mm]{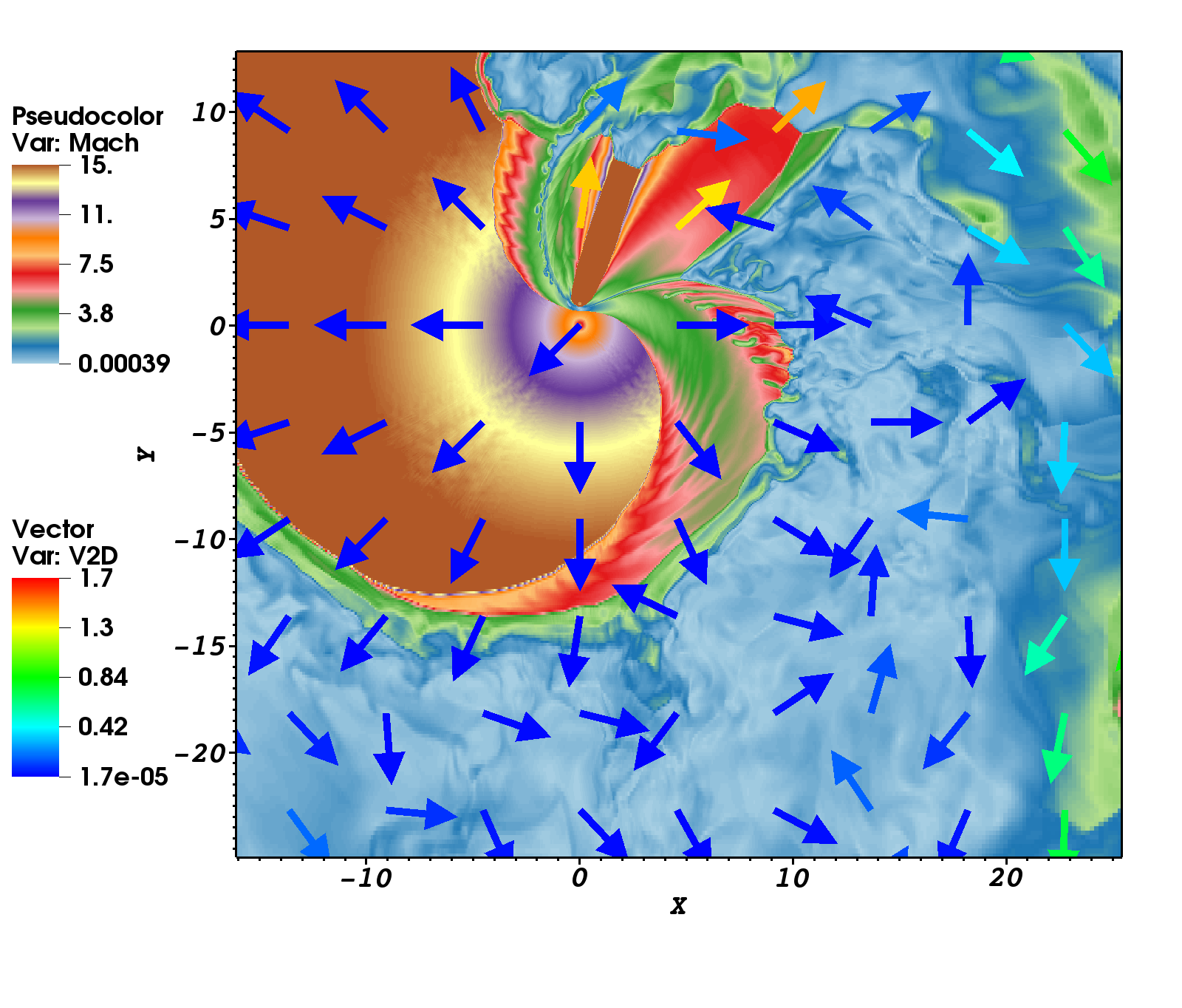}
\caption{Distribution of tracer, $\gamma\beta$, and Mach number (from top to bottom) for 2Dhbf at $t=5.2$~days.}
% ti3 model 
\label{fig:2dhbftum}
\end{figure}
%fffffffffffffffffffffffffffffffffffffffffffffffffffffffffffffffffff

%fffffffffffffffffffffffffffffffffffffffffffffffffffffffffffffffffff
\begin{figure*}[htp]
\centering
\includegraphics[width=80mm]{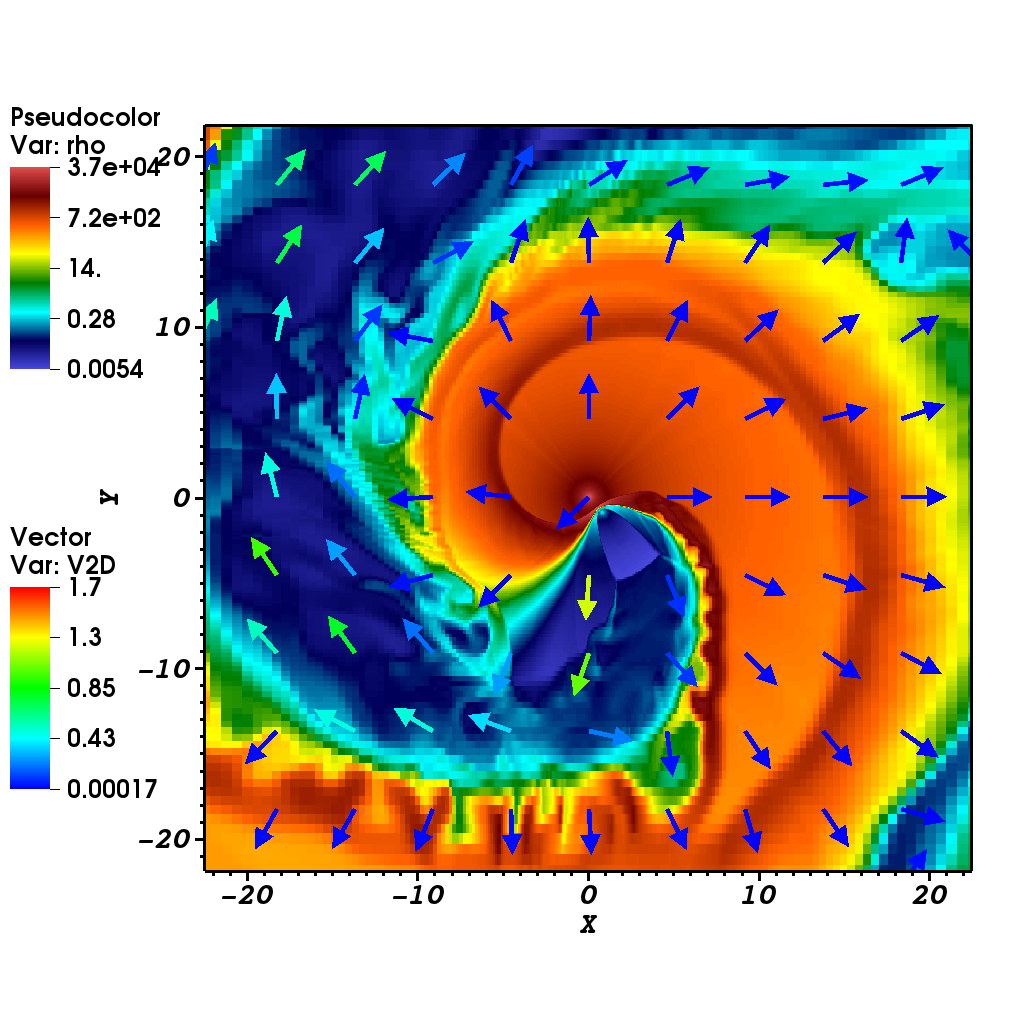}
\includegraphics[width=80mm]{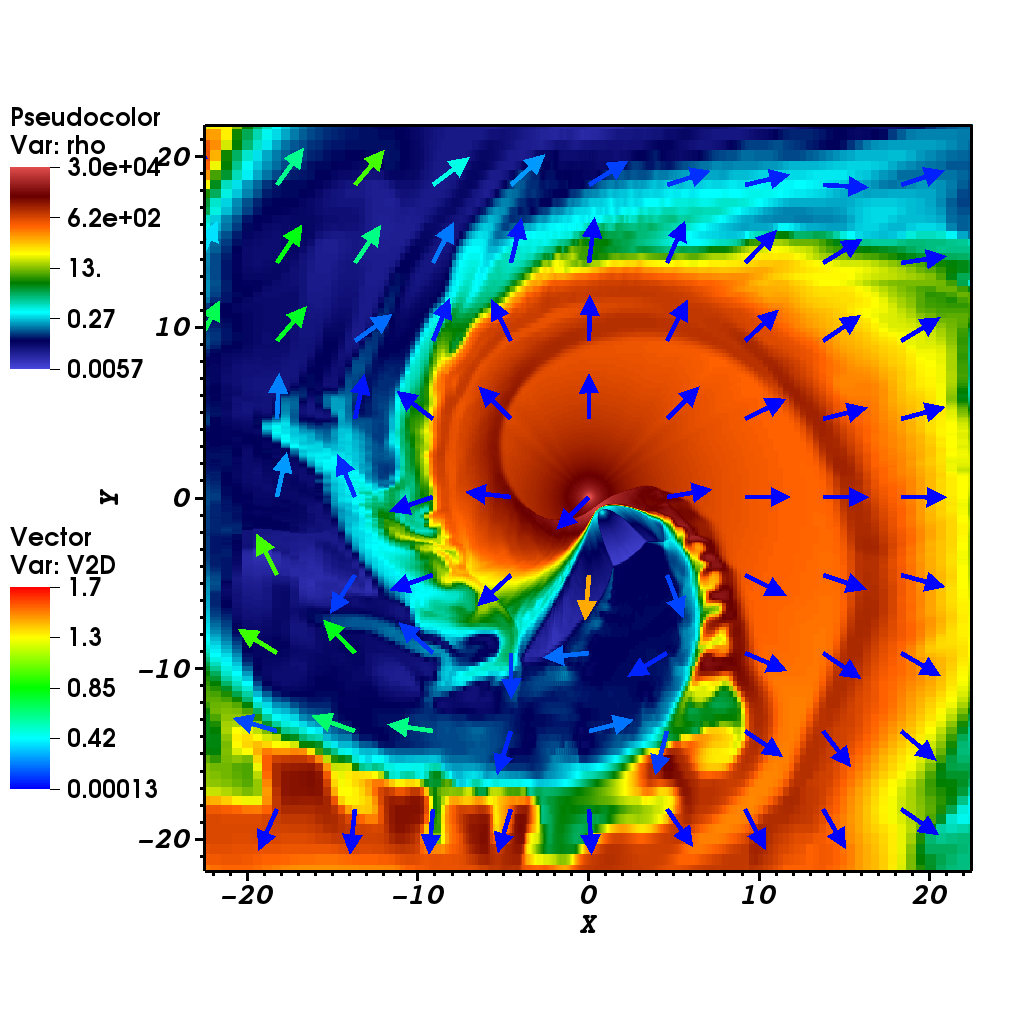}
\caption{Comparison of the density distribution at $t=3.1$~days between 2Dlf (left), and 2Dle (right), corresponding respectively to an ideal relativistic gas with constant $\hat\gamma=4/3$, and to an adaptive Synge-type EoS for a one particle species, non-degenerate, relativistic gas. The arrows represent the flow motion direction.}
% ti3 model 
\label{compeos}
\end{figure*}
%fffffffffffffffffffffffffffffffffffffffffffffffffffffffffffffffffff

\subsubsection{Evolution on larger scales}

The orbital motion leads to a spiral-like structure, as expected. However, the coherence of this structure is strongly affected by the unstable nature of the flow, 
which is rich in strong shocks and mixing induced by the Coriolis force, as predicted in \cite{bb11} and suggested by the numerical results of \cite{bbkp12}. 
We must note, however, that low resolution on the larger scales leads to the attenuation of shocks and turbulence and to the enhancement of mixing.

The present 3D results strengthen the thesis that the spiral 
shape is eventually lost, because the structure almost gets disrupted already within the grid, on scales $\sim 30\,a$. This is supported by 
Figs.~\ref{fig:3dsl}--\ref{fig:3dlfxzcph}, in which the lowest density regions of the {\it arm} of the spiral seem to split into different branches. 

The complexity of the matter distribution is best illustrated in the plane cuts perpendicular to the orbital plane in Figs.~\ref{fig:3dsl}, 
\ref{fig:3dlfxz}, and \ref{fig:3dlfxzcph}, which also show how the pulsar wind is terminated at moderately different distances from the pulsar 
depending on latitude. This is explained by the different pressure found by the pulsar wind, depending on the angle with which it faces the Coriolis force. 
The opening angle of the shocked-wind structure is determined by the Coriolis force on the orbital plane, whereas in the perpendicular direction,
the opening angle is $\sim 60^o$, as in the case without orbital motion \citep[e.g.][]{bog08}. However, the KHI  and RMI+RTI starting in the CD lead to fluctuations 
in the opening angle in this direction and, on larger scales, the shocked-wind structure becomes wider, since the high sound speed of the shocked gas allows the latter to quickly transfer
thrust to the vertical direction. 

We point out that in the vertical cuts what could be mostly distorting the CD should be KHI, because the RMI+RTI lack the impact of the Coriolis force. 
As seen in the perpendicular cuts of Figs.~\ref{fig:3dlfxz} and \ref{fig:3dlfxzcph}, variations in $d$ and in the pulsar angular velocity along the moderately eccentric orbit yield almost proportional variations in $R_{\rm p}$ and in the size of the whole interaction region. In particular, the ratio of $d$ between apastron and periastron passages is $\approx 1.8$, whereas the whole interaction region size changes by a factor $\approx 1.6$ in our simulations. Therefore, although size changes induced by instabilities are also taking place, as shown in Figs.~\ref{fig:3dlfxz} and \ref{fig:3dlfxzcph} and also in Fig.~\ref{fig:3Dlfxyoph}, the region size evolves correlated with the orbital motion.

%\subsubsection{Impact of stellar wind density and speed}

%With a density contrast three times larger, and a stellar wind velocity three times lower, the evolution of the shocked flows in 3Dls is qualitatively similar to that of 3Dlf (see %Fig.~\ref{3Dwcomp}). The shocked-wind structure shows the same features in both cases, but in 3Dls the spiral turns closer to the binary on the simulated scales as the stellar %wind-to-orbital velocity ratio is smaller \citep[see eq.~7 in][]{bb11}. 

%Larger density contrasts are to be explored through higher $\Gamma$-values, as the adopted approach leads to effects on the global geometry of the structure that may mask smaller %effects from the density contrast alone. Unfortunately, increasing the value of $\Gamma$ has very strong computational requirements in the a scenario including orbital motion. It is %important however to note that 2D simulations performed with $\Gamma=10$, accounting for orbital motion, and focused on the binary scales, showed qualitative agreement with the %results obtained for $\Gamma=2$, both regarding the generation of KHI, and the presence of a {\it Coriolis shock} \citep[see fig.~7 in][in which the Coriolis shock is also partially visible]{bbkp12}.

\subsection{Comparison of 2D and 3D results}

% %fffffffffffffffffffffffffffffffffffffffffffffffffffffffffffffffffff
% \begin{figure*}[h]
% \centering
% \includegraphics[width=80mm]{figures/2Dlf_rho_u4_xy_4p_t3150.png}
% \includegraphics[width=80mm]{figures/2Dlf_rho_u4_xy_4p_t4675.png}
% \includegraphics[width=80mm]{figures/2Dlf_rho_u4_xy_4p_t6200.png}
% \includegraphics[width=80mm]{figures/2Dlf_rho_u4_xy_4p_t7025.png}
% \caption{Density distribution by colour, and arrows representing the flow motion direction, for 2Dlf at: $t=2.6$, 3.9 (apastron), 5.2, 5.85~days (periastron) (from top to bottom, and left to right).}
% % ti3 model 
% \label{fig:2Dlfxyoph}
% \end{figure*}
% %fffffffffffffffffffffffffffffffffffffffffffffffffffffffffffffffffff

The results shown in Figs.~\ref{fig:2Dlfxyoph} and \ref{fig:2Dhbrho} are very similar to those obtained by \cite{bbkp12} from 2D simulations with slightly different orbital parameters, and are also qualitatively very much like the 3D results just presented. As noted, the main features of the quasi-stationary solution in 2D and 3D are the same. The density distribution maps of Figs.~\ref{fig:3Dlfxyoph} and \ref{fig:2Dlfxyoph} and in the top panel of Fig.~\ref{fig:comparison} allow for a comparison of results in 2D and 3D at periastron with the same resolution: the wind from the pulsar terminates closer to the pulsar, but it is wider in 3D than in 2D. Shocks, mixing, and turbulence start earlier, and mixing and turbulence 
seem to be much stronger in 3D. All this is also seen when comparing Figs.~\ref{fig:3dlftum} and \ref{fig:2dhbftum}.
%, although their resolution per dimension are not the same. 

% %fffffffffffffffffffffffffffffffffffffffffffffffffffffffffffffffffff
% \begin{figure*}[h]
% \centering
% \includegraphics[width=80mm]{figures/2Dhbf_rho_u4_xz_4p_t5800.png}
% \includegraphics[width=80mm]{figures/2Dhbf_rho_u4_xz_4p_t7550.png}
% \includegraphics[width=80mm]{figures/2Dhbf_rho_u4_xz_4p_t8450.png}
% \includegraphics[width=80mm]{figures/2Dhbf_rho_u4_xz_4p_t11825.png}
% \caption{Density distribution by colour, and arrows representing the flow motion direction, for 2Dhbf at: $t=4.8$, 6.3, 7.0, and 9.85~days ($\sim$ periastron) (from top to bottom, and left to right).}
% % ti3 model 
% \label{fig:2Dhbrho}
% \end{figure*}
% %fffffffffffffffffffffffffffffffffffffffffffffffffffffffffffffffffff

The different $\eta$-values adopted in 3D and 2D, 0.1 versus $0.1^{1/2}$ (see Sect.~\ref{num}), imply higher mass, momentum and energy fluxes for the pulsar wind in 2D, so it can explain why in 3D the termination of the pulsar wind and triggering of flow disruptive phenomena are closer to the pulsar. The higher strength of the disruptive phenomena in 3D would be, on the other hand, motivated by the higher number of geometric degrees of freedom. The wider opening angle of the unshocked pulsar wind region in 3D (see Fig.~\ref{fig:comparison}) could be explained by the stronger dilution of the shocked pulsar wind in the trailing edge of the interaction region.

\subsection{Effects of higher resolution}

% %fffffffffffffffffffffffffffffffffffffffffffffffffffffffffffffffffff
% \begin{figure}[h]
% \centering
% \includegraphics[width=70mm]{figures/2Dhbf_tr_u4_xy_t6200.png}
% \includegraphics[width=70mm]{figures/2Dhbf_mu4_u4_xy_t6200.png}
% \includegraphics[width=70mm]{figures/2Dhbf_mach_u4_xy_t6200.png}
% \caption{Distribution of tracer, $\gamma\beta$, and Mach number (from top to bottom) for 2Dhbf at $t=5.2$~days.}
% % ti3 model 
% \label{fig:2dhbftum}
% \end{figure}
% %fffffffffffffffffffffffffffffffffffffffffffffffffffffffffffffffffff

Figures~ \ref{fig:comparison}, \ref{fig:2Dlfxyoph}, and \ref{fig:2Dhbrho} (top right and bottom left and right panels) allow for a comparison of the impact of increasing the resolution on the simulation results. Qualitatively, the higher resolution does not change the shocked-wind structure significantly and, as when comparing 3D and 2D simulations, the same major features are present in all the simulations: (i) fast RMI+RTI and possibly KHI growth at the CD, (ii) strong shocks terminating both the unshocked and the shocked pulsar wind, (iii) bending of the whole structure, (iv) development of turbulence, (v) and strong two-wind mixing. However, with an increase in resolution, the instability growth is quicker, which leads to more turbulence downstream in the spiral and also to pulsar wind termination closer to the pulsar as turbulent material from the CD exerts more pressure from one side \citep[see][for a general study of the resolution influence on the evolution of KHI in numerical calculations]{perucho2004}. Given the higher dimensionality, a quicker instability development in 3D is expected, terminating the pulsar wind and generating turbulence even closer to the pulsar, if the resolution is increased. This must nevertheless be tested in future work with a higher resolution and pulsar-wind Lorentz factor.

\subsection{Effects of a larger grid}

Figure~\ref{fig:2Dhbrho} shows the density distribution evolution for the 2D simulation with the largest grid and the highest effective resolution. The maps show how the spiral starts disrupting through the growth of the KHI, RMI, and RTI, which favours two-wind mixing and structure deformation. In addition, different spiral arms are connected through channels of lower density, while the spiral arms themselves seem to lose their integrity. This 2D simulation strongly suggests that in 3D, the structure would lose coherence even closer ($<100\,a$) to the binary. The effects on these scales with a mildly relativistic pulsar wind in 2D are similar to those obtained by \cite{LambertsAAA} also in 2D for the non-relativistic case and specific regions of the parameter space.

\subsection{Effects of a different EoS}

Figure~\ref{compeos} allows for the comparison in 2D of using two different EoS, a single-species, ideal relativistic gas with constant $\hat\gamma=4/3$, and an adaptive Synge-type EoS for a one particle-species, non-degenerate, relativistic gas. Despite the different gas physics of the interacting flows, the figure shows that the results are almost identical in both cases, with only moderate changes in size for the inner region of the shocked-wind structure, in particular regarding the thickness of the shocked wind regions, which are thicker in the Taub case. Accordingly, the densities of the shocked flows also reach moderately lower values in the Taub case. All this is expected because the compression ratio is smaller for a non- ($\hat\gamma=5/3$) or a mildly ($\hat\gamma\in (4/3-5/3)$) relativistic wind than in the ultrarelativistic case, when $\hat\gamma=4/3$: from 4 to 7. This is also the reason for the {\it Coriolis shock} occurring slightly closer to the pulsar in the Taub case, since the shocked pulsar wind region is slightly thicker in that case.

\section{Discussion and summary}\label{disc}

% %fffffffffffffffffffffffffffffffffffffffffffffffffffffffffffffffffff
% \begin{figure*}[h]
% \centering
% \includegraphics[width=80mm]{figures/3Dlf_rho_u4_xz_t7025.png}
% \includegraphics[width=80mm]{figures/2Dlf_rho_u4_xz_t7025.png}
% \includegraphics[width=80mm]{figures/2Dhf_rho_u4_xz_t7025.png}
% \includegraphics[width=80mm]{figures/2Dhbf_rho_u4_xz_t7025.png}
% \caption{Comparison of the density distribution (colour), and flow motion direction (arrows), in the orbital plane (XY) at $t=5.85$~days (periastron) for 3Dlf, 2Dlf, 2Dhf, and 2Dhbf 
% (from top to bottom, and left to right).}
% % ti3 model 
% \label{fig:comparison}
% \end{figure*}
% %fffffffffffffffffffffffffffffffffffffffffffffffffffffffffffffffffff

The results presented here suggest that the shocked-wind structure is already very unstable on scales $\sim 10\,a$. In addition to structure bending, 
flow reacceleration, and full pulsar wind termination, all features found in both 2D and 3D simulations, the instabilities affecting the CD, possibly KHI and RMI+RTI, grow faster in 3D.
This leads to turbulence and to shocks further downstream in the outflowing shocked winds, and also to efficient matter mixing that leads to more shocks, 
because dense clumps of stellar material penetrate into the region occupied by the fast and light shocked pulsar wind. 

Most of the results of this work has already been found in 
previous 2D simulations \citep{bbkp12}, but in 3D all the processes occur closer to the binary. An enhancement of the resolution should make the 
interaction region more unsteady, likely shortening further the distances on which the RMI+RTI and KHI develop, and enhancing two-wind mixing even more. 
It is interesting to note that the bending length in 3D is about three times longer than the point of balance between the Coriolis force and the 
pulsar wind ram pressure \citep[see Eq.~9 in][]{bb11}. This is expected because the perturbations triggered by reaching pressure balance at the leading edge of the CD need time to transfer energy and momentum to the whole shocked-wind structure. 

On larger scales, $\sim 10-100\,a$, the radial pressure exerted by the shock mixed flow outwards triggers strong RTI in addition to KHI and RMI, which are already hinted at in 
3Dlf but are seen strongly developed in 2Dhbf, to the extent of disrupting and mixing the arms of the spiral formed by the orbital motion. 
This environment, which is still relatively close to the star to have enough radiation targets but also large enough to be resolvable in radio, 
is very rich in candidate sites for non-thermal emission (see the discussion in \citealt{bbkp12}; see also the semi-analytical study at high energies of \citealt{zbak13}) for which Doppler boosting can be an issue given 
the non-uniform velocity distribution. In particular, as seen for instance in Figs.~\ref{fig:3Dlfxyoph} and \ref{fig:2Dhbrho}, shocked pulsar 
wind material is found to move straight along quite large distances ($\sim 10\,a$). Since this material is shocked, it has a likely content of non-thermal particles.
Owing to the proximity to the star, these particles could quickly radiate their energy through inverse Compton scattering. This radiation, given the at least mildly relativistic velocities, could be Doppler boosted by a factor of $\sim 16\gamma^2\gtrsim 10$ for an observer viewing the flow with an angle $<1/\gamma$ from its axis, where $\gamma$ is the Lorentz factor. 
For large systems such as
PSR~B1259$-$63, X-rays could also be useful for observationally probing the shocked-wind structure on scales of $\sim 10^{17}$~cm \citep{kargaltsev2014}. For smaller systems and/or on larger scales \citep{durant2011}, one may study the bubble in the surrounding medium fed by the shocked two-wind flow \citep{bb11}.

It is noteworthy that, owing to momentum and energy isotropization, the evolution of the shocked-wind structure in the perpendicular direction gets broader further 
downstream than in the case without orbital motion. The study of this process deserves specific simulations with a larger grid in the $Z$ direction. 
In the same line, a more accurate characterization of the evolution of the shocked-wind structure (velocities, shocks, turbulence, mixing and general coherence) along the orbit would deserve higher resolution and eventually a larger grid size, despite being computationally very costly. 

Although our 3D simulations are capturing some relevant features of the shocked-wind structure, a natural future step should be to increase the grid resolution. For instance, to increase the Lorentz factor of the pulsar wind by a factor of a few, the resolution of the central part of the computational domain should be doubled. In addition, as hinted at by the 2D simulations (see Fig.~\ref{fig:comparison}), the large scale turbulence would be well resolved with three to four times more resolution in the grid than in our current 3D setup. This can be justified by a minimum cell size close to the typical gyroradius of the shocked pulsar wind particles: $\sim 3\times 10^9E_{\rm TeV}/B_{\rm G}$~cm, where $E_{\rm TeV}$ is the particle energy in TeV and $B_{\rm G}$ the magnetic field in G. This applies if one assumes that the physical dissipation scales are near the particle gyroradius, which is reasonable for a shocked pulsar wind consisting of an ideal collisionless plasma with low $B$ and with realistic $\Gamma$-values. However, such an increase in resolution would require about $10^7$~cpu~hours. This is a demanding calculation, but we plan to carry it out as it is a necessary step in the characterization of the scenario under study\footnote{We mean here average improvement, although in the grid periphery, the resolution would still be signicantly below the ideal case.}.
Another, less costly, improvement would be to use a more realistic EoS, although our preliminary study shows only minor differences between using CtGA and Taub. 
Thermal cooling should also be included since the shock of the stellar wind may be radiative, making the collision region more unstable and favouring mixing of the CD and its disruption \citep{pit09}. This is likely to increase turbulence and the presence of shocks in the shocked pulsar wind region even further. More feasible, but requiring the simplification of the region containing the system, is the simulation of the interaction of the shocked-wind structure with the environment on large scales, which was predicted to take place as a fast dense non-relativistic wind colliding with the surrounding medium \citep{bb11}. 

Another important line of research is computing  the emission from the shocked flows using the hydrodynamical information, which could be carried out first in the test-particle approximation. Another improvement would be to carry relativistic simulations with an anisotropic and/or inhomogeneous stellar wind\footnote{Regarding anisotropic and/or inhomogeneous (clumpy) stellar winds, see \cite{Paredes-Fortuny2015} for the axisymmetric case, and \cite{oka11,Takata2012} for the non-relativistic case.}, such as adding an equatorial disc, because several of the known gamma-ray binaries host Be stars. Finally, VLBI radio data are very important for observationally characterizing the shocked-wind structure on the simulated scales \citep[e.g.][]{moldon2011a,moldon2011b,moldon2012}.

\begin{acknowledgements} 
The calculations were carried out in the CFCA cluster of National Astronomical Observatory of Japan. We thank Andrea Mignone and the
{\it PLUTO} team for the possibility to use the {\it PLUTO} code. We especially thank to Claudio Zanni for technical support. 
The visualization of the results used the VisIt package. 
V.B-R. acknowledges support by the Spanish Ministerio de Econom\'{\i}a y Competitividad (MINECO) under grants AYA2013-47447-C3-1-P.
V.B-R. also acknowledges financial support from MINECO and European Social Funds through a Ram\'on y Cajal fellowship.
This research has been supported by the Marie Curie Career Integration Grant 321520. 
BMV acknowledges partial  support  by the JSPS (Japan Society for the Promotion of Science):
No.2503786, 25610056, 26287056, 26800159. BMV also acknowledges MEXT (Ministry of Education, Culture, Sports, Science and Technology):
No.26105521 and RFBR grant 12-02-01336-a.
MP is a member of the working team of projects AYA2013-40979-P and AYA2013-48226-C3-2-P, funded by the Spanish Ministerio de Economía y Competitividad.
\end{acknowledgements}

\bibliographystyle{aa}
%\bibliography{text,wc}
\bibliography{text}

\begin{thebibliography}{51}
\expandafter\ifx\csname natexlab\endcsname\relax\def\natexlab#1{#1}\fi

\bibitem[{{Abdo} {et~al.}(2011){Abdo}, {Ackermann}, {Ajello}, {Allafort},
  {Ballet}, {Barbiellini}, {Bastieri}, {Bechtol}, {Bellazzini}, {Berenji},
  {Blandford}, {Bonamente}, {Borgland}, {Bregeon}, {Brigida}, {Bruel},
  {Buehler}, {Buson}, {Caliandro}, {Cameron}, {Camilo}, {Caraveo}, {Cecchi},
  {Charles}, {Chaty}, {Chekhtman}, {Chernyakova}, {Cheung}, {Chiang},
  {Ciprini}, {Claus}, {Cohen-Tanugi}, {Cominsky}, {Corbel}, {Cutini},
  {D'Ammando}, {de Angelis}, {den Hartog}, {de Palma}, {Dermer}, {Digel},
  {Silva}, {Dormody}, {Drell}, {Drlica-Wagner}, {Dubois}, {Dubus}, {Dumora},
  {Enoto}, {Espinoza}, {Favuzzi}, {Fegan}, {Ferrara}, {Focke}, {Fortin},
  {Fukazawa}, {Funk}, {Fusco}, {Gargano}, {Gasparrini}, {Gehrels}, {Germani},
  {Giglietto}, {Giommi}, {Giordano}, {Giroletti}, {Glanzman}, {Godfrey},
  {Grenier}, {Grondin}, {Grove}, {Grundstrom}, {Guiriec}, {Gwon}, {Hadasch},
  {Harding}, {Hayashida}, {Hays}, {J{\'o}hannesson}, {Johnson}, {Johnson},
  {Johnston}, {Kamae}, {Katagiri}, {Kataoka}, {Keith}, {Kerr},
  {Kn{\"o}dlseder}, {Kramer}, {Kuss}, {Lande}, {Lee}, {Lemoine-Goumard},
  {Longo}, {Loparco}, {Lovellette}, {Lubrano}, {Manchester}, {Marelli},
  {Mazziotta}, {Michelson}, {Mitthumsiri}, {Mizuno}, {Moiseev}, {Monte},
  {Monzani}, {Morselli}, {Moskalenko}, {Murgia}, {Nakamori}, {Naumann-Godo},
  {Neronov}, {Nolan}, {Norris}, {Noutsos}, {Nuss}, {Ohsugi}, {Okumura},
  {Omodei}, {Orlando}, {Paneque}, {Parent}, {Pesce-Rollins}, {Pierbattista},
  {Piron}, {Porter}, {Possenti}, {Rain{\`o}}, {Rando}, {Ray}, {Razzano},
  {Razzaque}, {Reimer}, {Reimer}, {Reposeur}, {Ritz}, {Sadrozinski}, {Scargle},
  {Sgr{\`o}}, {Shannon}, {Siskind}, {Smith}, {Spandre}, {Spinelli},
  {Strickman}, {Suson}, {Takahashi}, {Tanaka}, {Thayer}, {Thayer}, {Thompson},
  {Thorsett}, {Tibaldo}, {Tibolla}, {Torres}, {Tosti}, {Troja}, {Uchiyama},
  {Usher}, {Vandenbroucke}, {Vasileiou}, {Vianello}, {Vitale}, {Waite}, {Wang},
  {Winer}, {Wolff}, {Wood}, {Wood}, {Yang}, {Ziegler}, \& {Zimmer}}]{abd11}
{Abdo}, A.~A., {Ackermann}, M., {Ajello}, M., {et~al.} 2011, \apjl, 736, L11

\bibitem[{{Aharonian} {et~al.}(2005){Aharonian}, {Akhperjanian}, {Aye},
  {Bazer-Bachi}, {Beilicke}, {Benbow}, {Berge}, {Berghaus}, {Bernl{\"o}hr},
  {Boisson}, {Bolz}, {Braun}, {Breitling}, {Brown}, {Bussons Gordo},
  {Chadwick}, {Chounet}, {Cornils}, {Costamante}, {Degrange},
  {Djannati-Ata{\"i}}, {O'C.~Drury}, {Dubus}, {Emmanoulopoulos}, {Espigat},
  {Feinstein}, {Fleury}, {Fontaine}, {Fuchs}, {Funk}, {Gallant}, {Giebels},
  {Gillessen}, {Glicenstein}, {Goret}, {Hadjichristidis}, {Hauser},
  {Heinzelmann}, {Henri}, {Hermann}, {Hinton}, {Hofmann}, {Holleran}, {Horns},
  {de Jager}, {Johnston}, {Kh{\'e}lifi}, {Kirk}, {Komin}, {Konopelko},
  {Latham}, {Le Gallou}, {Lemi{\`e}re}, {Lemoine-Goumard}, {Leroy},
  {Martineau-Huynh}, {Lohse}, {Marcowith}, {Masterson}, {McComb}, {de Naurois},
  {Nolan}, {Noutsos}, {Orford}, {Osborne}, {Ouchrif}, {Panter}, {Pelletier},
  {Pita}, {P{\"u}hlhofer}, {Punch}, {Raubenheimer}, {Raue}, {Raux}, {Rayner},
  {Redondo}, {Reimer}, {Reimer}, {Ripken}, {Rob}, {Rolland}, {Rowell},
  {Sahakian}, {Saug{\'e}}, {Schlenker}, {Schlickeiser}, {Schuster}, {Schwanke},
  {Siewert}, {Skj{\ae}raasen}, {Sol}, {Steenkamp}, {Stegmann}, {Tavernet},
  {Terrier}, {Th{\'e}oret}, {Tluczykont}, {Vasileiadis}, {Venter}, {Vincent},
  {V{\"o}lk}, \& {Wagner}}]{aha05}
{Aharonian}, F., {Akhperjanian}, A.~G., {Aye}, K.-M., {et~al.} 2005, \aap, 442,
  1

\bibitem[{{Aharonian} {et~al.}(2012){Aharonian}, {Bogovalov}, \&
  {Khangulyan}}]{abk12}
{Aharonian}, F.~A., {Bogovalov}, S.~V., \& {Khangulyan}, D. 2012, \nat, 482,
  507

\bibitem[{{Aragona} {et~al.}(2009){Aragona}, {McSwain}, {Grundstrom}, {Marsh},
  {Roettenbacher}, {Hessler}, {Boyajian}, \& {Ray}}]{Aragona2009}
{Aragona}, C., {McSwain}, M.~V., {Grundstrom}, E.~D., {et~al.} 2009, \apj, 698,
  514

\bibitem[{{Barkov} \& {Khangulyan}(2012)}]{barkha12}
{Barkov}, M.~V. \& {Khangulyan}, D.~V. 2012, \mnras, 2375

\bibitem[{{Bogovalov} {et~al.}(2012){Bogovalov}, {Khangulyan}, {Koldoba},
  {Ustyugova}, \& {Aharonian}}]{bog12}
{Bogovalov}, S.~V., {Khangulyan}, D., {Koldoba}, A.~V., {Ustyugova}, G.~V., \&
  {Aharonian}, F.~A. 2012, \mnras, 419, 3426

\bibitem[{{Bogovalov} {et~al.}(2008){Bogovalov}, {Khangulyan}, {Koldoba},
  {Ustyugova}, \& {Aharonian}}]{bog08}
{Bogovalov}, S.~V., {Khangulyan}, D.~V., {Koldoba}, A.~V., {Ustyugova}, G.~V.,
  \& {Aharonian}, F.~A. 2008, \mnras, 387, 63

\bibitem[{{Bosch-Ramon} \& {Barkov}(2011)}]{bb11}
{Bosch-Ramon}, V. \& {Barkov}, M.~V. 2011, \aap, 535, A20

\bibitem[{{Bosch-Ramon} {et~al.}(2012){Bosch-Ramon}, {Barkov}, {Khangulyan}, \&
  {Perucho}}]{bbkp12}
{Bosch-Ramon}, V., {Barkov}, M.~V., {Khangulyan}, D., \& {Perucho}, M. 2012,
  \aap, 544, A59

\bibitem[{{Brouillette}(2002)}]{b02}
{Brouillette}, M. 2002, Annual Review of Fluid Mechanics, 34, 445

\bibitem[{{Bucciantini} {et~al.}(2005){Bucciantini}, {Amato}, \& {Del
  Zanna}}]{bucciantini2005}
{Bucciantini}, N., {Amato}, E., \& {Del Zanna}, L. 2005, \aap, 434, 189

\bibitem[{{Casares} {et~al.}(2005){Casares}, {Rib{\'o}}, {Ribas}, {Paredes},
  {Mart{\'{\i}}}, \& {Herrero}}]{Casares2005}
{Casares}, J., {Rib{\'o}}, M., {Ribas}, I., {et~al.} 2005, \mnras, 364, 899

\bibitem[{{Chandrasekhar}(1961)}]{cha61}
{Chandrasekhar}, S. 1961, {Hydrodynamic and hydromagnetic stability}

\bibitem[{{Chernyakova} {et~al.}(2014){Chernyakova}, {Abdo}, {Neronov},
  {McSwain}, {Mold{\'o}n}, {Rib{\'o}}, {Paredes}, {Sushch}, {de Naurois},
  {Schwanke}, {Uchiyama}, {Wood}, {Johnston}, {Chaty}, {Coleiro}, {Malyshev},
  \& {Babyk}}]{Chernyakova2014}
{Chernyakova}, M., {Abdo}, A.~A., {Neronov}, A., {et~al.} 2014, \mnras, 439,
  432

\bibitem[{{Colella} \& {Woodward}(1984)}]{cw84}
{Colella}, P. \& {Woodward}, P.~R. 1984, Journal of Computational Physics, 54,
  174

\bibitem[{{Dubus}(2013)}]{Dubus2013}
{Dubus}, G. 2013, \aapr, 21, 64

\bibitem[{{Durant} {et~al.}(2011){Durant}, {Kargaltsev}, {Pavlov}, {Chang}, \&
  {Garmire}}]{durant2011}
{Durant}, M., {Kargaltsev}, O., {Pavlov}, G.~G., {Chang}, C., \& {Garmire},
  G.~P. 2011, \apj, 735, 58

\bibitem[{{Inoue}(2012)}]{i12}
{Inoue}, T. 2012, \apj, 760, 43

\bibitem[{{Johnston} {et~al.}(1992){Johnston}, {Manchester}, {Lyne}, {Bailes},
  {Kaspi}, {Qiao}, \& {D'Amico}}]{joh92}
{Johnston}, S., {Manchester}, R.~N., {Lyne}, A.~G., {et~al.} 1992, \apjl, 387,
  L37

\bibitem[{{Kargaltsev} {et~al.}(2014){Kargaltsev}, {Pavlov}, {Durant},
  {Volkov}, \& {Hare}}]{kargaltsev2014}
{Kargaltsev}, O., {Pavlov}, G.~G., {Durant}, M., {Volkov}, I., \& {Hare}, J.
  2014, \apj, 784, 124

\bibitem[{{Kennel} \& {Coroniti}(1984)}]{ken84}
{Kennel}, C.~F. \& {Coroniti}, F.~V. 1984, \apj, 283, 710

\bibitem[{{Khangulyan} {et~al.}(2012){Khangulyan}, {Aharonian}, {Bogovalov}, \&
  {Ribo}}]{kha12}
{Khangulyan}, D., {Aharonian}, F.~A., {Bogovalov}, S.~V., \& {Ribo}, M. 2012,
  \apj, in press

\bibitem[{{Lamberts} {et~al.}(2012{\natexlab{a}}){Lamberts}, {Dubus},
  {Fromang}, \& {Lesur}}]{LambertsBBB}
{Lamberts}, A., {Dubus}, G., {Fromang}, S., \& {Lesur}, G. 2012{\natexlab{a}},
  in American Institute of Physics Conference Series, Vol. 1505, American
  Institute of Physics Conference Series, ed. F.~A. {Aharonian}, W.~{Hofmann},
  \& F.~M. {Rieger}, 406--409

\bibitem[{{Lamberts} {et~al.}(2012{\natexlab{b}}){Lamberts}, {Dubus}, {Lesur},
  \& {Fromang}}]{LambertsAAA}
{Lamberts}, A., {Dubus}, G., {Lesur}, G., \& {Fromang}, S. 2012{\natexlab{b}},
  \aap, 546, A60

\bibitem[{{Lamberts} {et~al.}(2011){Lamberts}, {Fromang}, \&
  {Dubus}}]{lamberts2011}
{Lamberts}, A., {Fromang}, S., \& {Dubus}, G. 2011, \mnras, 418, 2618

\bibitem[{{Lamberts} {et~al.}(2013){Lamberts}, {Fromang}, {Dubus}, \&
  {Teyssier}}]{Lamberts2013}
{Lamberts}, A., {Fromang}, S., {Dubus}, G., \& {Teyssier}, R. 2013, \aap, 560,
  A79

\bibitem[{{Maraschi} \& {Treves}(1981)}]{Maraschi1981}
{Maraschi}, L. \& {Treves}, A. 1981, \mnras, 194, 1P

\bibitem[{{Matsumoto} \& {Masada}(2013)}]{mm13}
{Matsumoto}, J. \& {Masada}, Y. 2013, \apjl, 772, L1

\bibitem[{Meshkov(1969)}]{m69}
Meshkov, E. 1969, Fluid Dynamics, 4, 101

\bibitem[{{Mignone} \& {Bodo}(2005)}]{mig05}
{Mignone}, A. \& {Bodo}, G. 2005, \mnras, 364, 126

\bibitem[{{Mignone} {et~al.}(2007){Mignone}, {Bodo}, {Massaglia}, {Matsakos},
  {Tesileanu}, {Zanni}, \& {Ferrari}}]{mbm07}
{Mignone}, A., {Bodo}, G., {Massaglia}, S., {et~al.} 2007, \apjs, 170, 228

\bibitem[{{Mignone} {et~al.}(2005){Mignone}, {Plewa}, \& {Bodo}}]{mpb05}
{Mignone}, A., {Plewa}, T., \& {Bodo}, G. 2005, \apjs, 160, 199

\bibitem[{{Mold{\'o}n} {et~al.}(2011{\natexlab{a}}){Mold{\'o}n}, {Johnston},
  {Rib{\'o}}, {Paredes}, \& {Deller}}]{moldon2011b}
{Mold{\'o}n}, J., {Johnston}, S., {Rib{\'o}}, M., {Paredes}, J.~M., \&
  {Deller}, A.~T. 2011{\natexlab{a}}, \apjl, 732, L10

\bibitem[{{Mold{\'o}n} {et~al.}(2011{\natexlab{b}}){Mold{\'o}n}, {Rib{\'o}}, \&
  {Paredes}}]{moldon2011a}
{Mold{\'o}n}, J., {Rib{\'o}}, M., \& {Paredes}, J.~M. 2011{\natexlab{b}}, \aap,
  533, L7

\bibitem[{{Mold{\'o}n} {et~al.}(2012){Mold{\'o}n}, {Rib{\'o}}, \&
  {Paredes}}]{moldon2012}
---. 2012, \aap, 548, A103

\bibitem[{{Negueruela} {et~al.}(2011){Negueruela}, {Rib{\'o}}, {Herrero},
  {Lorenzo}, {Khangulyan}, \& {Aharonian}}]{neg11}
{Negueruela}, I., {Rib{\'o}}, M., {Herrero}, A., {et~al.} 2011, \apjl, 732, L11

\bibitem[{{Nishihara} {et~al.}(2010){Nishihara}, {Wouchuk}, {Matsuoka},
  {Ishizaki}, \& {Zhakhovsky}}]{nwm10}
{Nishihara}, K., {Wouchuk}, J.~G., {Matsuoka}, C., {Ishizaki}, R., \&
  {Zhakhovsky}, V.~V. 2010, Royal Society of London Philosophical Transactions
  Series A, 368, 1769

\bibitem[{{Okazaki} {et~al.}(2011){Okazaki}, {Nagataki}, {Naito}, {Kawachi},
  {Hayasaki}, {Owocki}, \& {Takata}}]{oka11}
{Okazaki}, A.~T., {Nagataki}, S., {Naito}, T., {et~al.} 2011, \pasj, 63, 893

\bibitem[{{Paredes} {et~al.}(2013){Paredes}, {Bednarek}, {Bordas},
  {Bosch-Ramon}, {De Cea del Pozo}, {Dubus}, {Funk}, {Hadasch}, {Khangulyan},
  {Markoff}, {Mold{\'o}n}, {Munar-Adrover}, {Nagataki}, {Naito}, {de Naurois},
  {Pedaletti}, {Reimer}, {Rib{\'o}}, {Szostek}, {Terada}, {Torres}, {Zabalza},
  {Zdziarski}, \& {CTA Consortium}}]{Paredes2013}
{Paredes}, J.~M., {Bednarek}, W., {Bordas}, P., {et~al.} 2013, Astroparticle
  Physics, 43, 301

\bibitem[{{Paredes-Fortuny} {et~al.}(2015){Paredes-Fortuny}, {Bosch-Ramon},
  {Perucho}, \& {Rib{\'o}}}]{Paredes-Fortuny2015}
{Paredes-Fortuny}, X., {Bosch-Ramon}, V., {Perucho}, M., \& {Rib{\'o}}, M.
  2015, \aap, 574, A77

\bibitem[{{Perucho} {et~al.}(2004){Perucho}, {Hanasz}, {Mart{\'{\i}}}, \&
  {Sol}}]{perucho2004}
{Perucho}, M., {Hanasz}, M., {Mart{\'{\i}}}, J.~M., \& {Sol}, H. 2004, \aap,
  427, 415

\bibitem[{{Pittard}(2009)}]{pit09}
{Pittard}, J.~M. 2009, \mnras, 396, 1743

\bibitem[{{Portegies Zwart} \& {Yungelson}(1999)}]{1999MNRAS.309...26P}
{Portegies Zwart}, S.~F. \& {Yungelson}, L.~R. 1999, \mnras, 309, 26

\bibitem[{Richtmyer(1960)}]{richt60}
Richtmyer, R.~D. 1960, Communications on Pure and Applied Mathematics, 13, 297

\bibitem[{{Romero} {et~al.}(2007){Romero}, {Okazaki}, {Orellana}, \&
  {Owocki}}]{Romero2007}
{Romero}, G.~E., {Okazaki}, A.~T., {Orellana}, M., \& {Owocki}, S.~P. 2007,
  \aap, 474, 15

\bibitem[{{Sarty} {et~al.}(2011){Sarty}, {Szalai}, {Kiss}, {Matthews}, {Wu},
  {Kuschnig}, {Guenther}, {Moffat}, {Rucinski}, {Sasselov}, {Weiss}, {Huziak},
  {Johnston}, {Phillips}, \& {Ashley}}]{Sarty2011}
{Sarty}, G.~E., {Szalai}, T., {Kiss}, L.~L., {et~al.} 2011, \mnras, 411, 1293

\bibitem[{{Takata} {et~al.}(2012){Takata}, {Okazaki}, {Nagataki}, {Naito},
  {Kawachi}, {Lee}, {Mori}, {Hayasaki}, {Yamaguchi}, \& {Owocki}}]{Takata2012}
{Takata}, J., {Okazaki}, A.~T., {Nagataki}, S., {et~al.} 2012, \apj, 750, 70

\bibitem[{{Tam} {et~al.}(2011){Tam}, {Huang}, {Takata}, {Hui}, {Kong}, \&
  {Cheng}}]{tam11}
{Tam}, P.~H.~T., {Huang}, R.~H.~H., {Takata}, J., {et~al.} 2011, \apjl, 736,
  L10

\bibitem[{{Taub}(1978)}]{taub78}
{Taub}, A.~H. 1978, Annual Review of Fluid Mechanics, 10, 301

\bibitem[{{Taylor}(1950)}]{t50}
{Taylor}, G. 1950, Royal Society of London Proceedings Series A, 201, 192

\bibitem[{{Zabalza} {et~al.}(2013){Zabalza}, {Bosch-Ramon}, {Aharonian}, \&
  {Khangulyan}}]{zbak13}
{Zabalza}, V., {Bosch-Ramon}, V., {Aharonian}, F., \& {Khangulyan}, D. 2013,
  \aap, 551, A17

\end{thebibliography}

\end{document}